\documentclass[longauth]{aa}  

\usepackage{graphicx}
\usepackage{color}
\usepackage{txfonts}
\usepackage{tablefootnote}
\begin{document} 

\makeatletter
\def\instrefs#1{{\def\scsep{\def\scsep{,}}\@for\w:=#1\do{\scsep\ref{inst:\w}}}}
\renewcommand{\inst}[1]{\unskip$^{\instrefs{#1}}$}
\newcommand{\akihiko}{\textcolor{blue}}

\title{TOI-1442 b and TOI-2445 b: two potentially rocky ultra-short period planets around M dwarfs}

\author{G. Morello\inst{iac,ull,inaf_palermo,chalmers}
\and
H. Parviainen\inst{iac,ull}
\and
F. Murgas\inst{iac,ull}
\and
E. Pall{\'e}\inst{iac,ull}
\and
M. Oshagh\inst{iac,ull}
\and
A. Fukui\inst{komaba,iac}
\and
T. Hirano\inst{abc,naoj}
\and
H. T. Ishikawa\inst{abc,naoj}
\and
M. Mori\inst{da_gsc}
\and
N. Narita\inst{komaba,abc,iac}
\and
K. A. Collins\inst{harvard}
\and
K. Barkaoui\inst{uliege1,mit,iac}
\and
P. Lewin\inst{maury_lewin}
\and
C. Cadieux\inst{omm}
\and
J. P. de Leon\inst{da_gsc}
\and
A. Soubkiou\inst{cadi-ayyad,porto1,porto2}
\and
N. Abreu Garcia\inst{iac,ull}
\and
N. Crouzet\inst{estec}
\and
E. Esparza-Borges\inst{iac,ull}
\and
G. E. Fern{\'a}ndez Rodr{\'i}guez\inst{iac}
\and
D. Galán\inst{iac,ull}
\and
Y. Hori\inst{abc,naoj}
\and
M. Ikoma\inst{ds_naoj}
\and
K. Isogai\inst{okayama,dms_gsas}
\and
T. Kagetani\inst{dms_gsas}
\and
K. Kawauchi\inst{iac,ull,komaba}
\and
T. Kimura\inst{deps_gsc}
\and
T. Kodama\inst{komaba}
\and
J. Korth\inst{gothen,lund1,lund2}
\and
T. Kotani\inst{abc,naoj,das_guas}
\and
V. Krishnamurthy\inst{abc,naoj}
\and
S. Kurita\inst{deps_gsc}
\and
A. Laza-Ramos\inst{iac}
\and
J. H. Livingston\inst{da_gsc}
\and
R. Luque\inst{iaa}
\and
A. Madrigal-Aguado\inst{iac,ull}
\and
T. Nishiumi\inst{das_guas,abc,dms_gsas}
\and
J. Orell-Miquel\inst{iac,ull}
\and
M. Puig-Subir{\`a}\inst{iac,ull}
\and
M. S{\'a}nchez-Benavente\inst{iac,ull}
\and
M. Stangret\inst{inaf_padova,iac,ull}
\and
M. Tamura\inst{da_gsc,abc,naoj}
\and
Y. Terada\inst{iaaas,dantu}
\and
N. Watanabe\inst{dms_gsas}
\and
Y. Zou\inst{dms_gsas}
\and
Z. Benkhaldoun\inst{cadi-ayyad}
\and
K. I. Collins\inst{george_mason}
\and
R. Doyon\inst{omm2}
\and
L. Garcia\inst{uliege1}
\and
M. Ghachoui\inst{uliege1,cadi-ayyad}
\and
M. Gillon\inst{uliege1}
\and
E. Jehin\inst{uliege2}
\and
F. J. Pozuelos\inst{iaa,uliege1}
\and
R. P. Schwarz\inst{harvard}
\and
M. Timmermans\inst{uliege1}
}

\institute{
\label{inst:iac}Instituto de Astrof\'isica de Canarias (IAC), 38205 La Laguna, Tenerife, Spain
\and 
\label{inst:ull}Departamento de Astrof\'isica, Universidad de La Laguna (ULL), 38206, La Laguna, Tenerife, Spain
\and
\label{inst:inaf_palermo}INAF- Palermo Astronomical Observatory, Piazza del Parlamento, 1, 90134 Palermo, Italy
\and
\label{inst:chalmers}Department of Space, Earth and Environment, Chalmers University of Technology, SE-412 96 Gothenburg, Sweden
\and
\label{inst:inaf_padova}INAF – Osservatorio Astronomico di Padova, Vicolo dell'Osservatorio 5, 35122, Padova, Italy
\and
\label{inst:komaba}Komaba Institute for Science, The University of Tokyo, 3-8-1 Komaba, Meguro, Tokyo 153-8902, Japan
\and
\label{inst:abc}Astrobiology Center, 2-21-1 Osawa, Mitaka, Tokyo 181-8588, Japan
\and
\label{inst:harvard}Center for Astrophysics \textbar \ Harvard \&
Smithsonian, 60 Garden Street, Cambridge, MA 02138, USA
\and
\label{inst:maury_lewin}The Maury Lewin Astronomical Observatory,
Glendora, California, 91741, USA
\and
\label{inst:omm}Universit\'e de Montr\'eal, D\'epartement de Physique, IREX, Montr\'eal, QC H3C 3J7, Canada
\and
\label{inst:omm2}Observatoire du Mont-M\'egantic, Universit\'e de Montr\'eal, Montr\'eal, QC H3C 3J7, Canada
\and
\label{inst:estec}European Space Agency (ESA), European Space Research and Technology Centre (ESTEC), Keplerlaan 1, 2201 AZ Noordwijk, The Netherlands
\and
\label{inst:naoj}National Astronomical Observatory of Japan, 2-21-1 Osawa, Mitaka, Tokyo 181-8588, Japan
\and
\label{inst:da_gsc}Department of Astronomy, Graduate School of Science, The University of Tokyo, 7-3-1 Hongo, Bunkyo-ku, Tokyo 113-0033, Japan
\and
\label{inst:okayama}Okayama Observatory, Kyoto University, 3037-5 Honjo, Kamogatacho, Asakuchi, Okayama 719-0232, Japan
\and
\label{inst:dms_gsas}Department of Multi-Disciplinary Sciences, Graduate School of Arts and Sciences, The University of Tokyo, 3-8-1 Komaba, Meguro, Tokyo 153-8902, Japan
\and
\label{inst:das_guas}Department of Astronomical Science, The Graduated University for Advanced Studies, SOKENDAI, 2-21-1, Osawa, Mitaka, Tokyo, 181-8588, Japan
\and
\label{inst:gothen}Department of Space, Earth and Environment, Astronomy and Plasma Physics, Chalmers University of Technology, 412 96 Gothenburg, Sweden
\and
\label{inst:lund1}Division of Astrophysics, Department of Physics, Lund University, Box 43, 22100 Lund, Sweden
\and
\label{inst:lund2}Lund Observatory, Department of Astronomy and Theoretical Physics, Lund University, Box 43, SE-221 00 Lund, Sweden
\and
\label{inst:iaa}Instituto de Astrof{\'i}sica de Andaluc{\'i}a (IAA-CSIC), Glorieta de la Astronom{\'i}a s/n, 18008 Granada, Spain
\and
\label{inst:iaaas}Institute of Astronomy and Astrophysics, Academia Sinica, P.O. Box 23-141, Taipei 10617, Taiwan, R.O.C.
\and
\label{inst:dantu}Department of Astrophysics, National Taiwan University, Taipei 10617, Taiwan, R.O.C.
\and
\label{inst:deps_gsc}Department of Earth and Planetary Science, Graduate School of Science, The University of Tokyo, 7-3-1 Hongo, Bunkyo-ku, Tokyo 113-0033, Japan
\and
\label{inst:ds_naoj}National Astronomical Observatory of Japan, 2-21-1 Osawa, Mitaka, Tokyo 181-8588, Japan
\and
\label{inst:uliege1}Astrobiology Research Unit, Universit\'e de Li\`ege, 19C All\'ee du 6 Ao\^ut, 4000 Li\`ege, Belgium
\and
\label{inst:mit}Department of Earth, Atmospheric and Planetary Science, Massachusetts Institute of Technology, 77 Massachusetts Avenue, Cambridge, MA 02139, USA
\and
\label{inst:uliege2}Space sciences, Technologies and Astrophysics Research (STAR) Institute, Universit\'e de Li\`ege, Belgium
\and
\label{inst:cadi-ayyad}Oukaimeden Observatory, High Energy Physics and Astrophysics Laboratory, Cadi Ayyad University, Marrakech, Morocco
\and
\label{inst:porto1}Departamento de Fisica e Astronomia, Faculdade de Ciencias, Universidade do Porto, Rua do Campo Alegre, 4169-007 Porto, Portugal
\and
\label{inst:porto2}Instituto de Astrofisica e Ciencias do Espaco, Universidade do porto, CAUP, Rua das Estrelas, 150-762 Porto, Portugal
\and
\label{inst:george_mason}George Mason University, 4400 University Drive, Fairfax, VA, 22030 USA
}


 
  \abstract
   {Exoplanets with orbital periods of less than one day are known as ultra-short period (USP) planets. They are relatively rare products of planetary formation and evolution processes, but especially favourable for characterisation with current planet detection methods. At the time of writing, 125 USP planets have already been confirmed.}
   {Our aim is to validate the planetary nature of two new transiting planet candidates around M dwarfs announced by the NASA Transiting Exoplanet Survey Satellite (TESS), registered as TESS Objects of Interest (TOIs) TOI-1442.01 and TOI-2445.01.}
   {We used   TESS data, ground-based photometric light curves, and Subaru/IRD spectrograph radial velocity (RV) measurements to validate both planetary candidates and to establish their physical properties.}
   {TOI-1442 b is a validated exoplanet with an orbital period of $P = 0.4090682 \pm 0.0000004 \, d$, a radius of $R_{\mathrm{p}} = 1.15 \pm 0.06 \, R_{\oplus}$, and equilibrium temperature of $T_{\mathrm{p,eq}} = 1357_{-42}^{+49} \, K$. TOI-2445 b is also validated with an orbital period of $P = 0.3711286 \pm 0.0000004 \, d$, a radius of $R_{\mathrm{p}} = 1.33 \pm 0.09 \, R_{\oplus}$, and equilibrium temperature of $T_{\mathrm{p,eq}} = 1330_{-56}^{+61} \, K$. Their physical properties align with current empirical trends and formation theories of USP planets. Based on the RV measurements, we set 3$\sigma$ upper mass limits of 8 $M_{\oplus}$ and 20 $M_{\oplus}$, thus confirming the non-stellar, sub-Jovian nature of both transiting objects. More RV measurements will be needed to constrain the planetary masses and mean densities, and the predicted presence of outer planetary companions. These targets extend the small sample of USP planets orbiting around M dwarfs up to 21 members. They are also among the 20 most suitable terrestrial planets for atmospheric characterisation via secondary eclipse with the James Webb Space Telescope, according to a widespread emission spectroscopy metric.}
   {}

\keywords{planetary systems --
                planets and satellites: individual: TOI-1442 b, TOI-2445 b --
                techniques: photometric --
                techniques: spectroscopic --
                methods: observational
               }

\maketitle
%

\section{Introduction}
The main theories of planetary formation based on our Solar System did not predict the existence of planets on orbits much narrower than that of Mercury (e.g. \citealp{lissauer1993,lin1996,bodenheimer2000}). However, since the discovery of the first hot Jupiter \citep{mayor1995}, many exoplanets have been detected with orbital periods of just a few days. Among the close-in planet population, an ultra-short period (USP) planet is defined as one that completes its entire orbit in less than one day \citep{sahu2006}. These extreme cases may provide some of the most revealing insights into the formation and evolution processes of planetary systems \citep{winn2018}. For example, USP planets are important to constrain the slope of the radius valley of close-in small transiting planets (e.g. \citealp{fulton2017,gupta2022,luque2022}). They are also excellent test beds for studying star--planet interactions, such as the effects of tidal forces and atmospheric erosion processes (e.g. \citealp{lopez2017,hamer2020,alvarado-montes2021}). 

All other conditions being equal, USP planets are the easiest to detect and characterise by means of occultations and radial velocity (RV) measurements \citep{gaudi2005}. Despite this strong selection bias, the observed period distribution of exoplanets drops at $P \lesssim 4 \, \mathrm{d}$, indicating that planets with shorter orbital periods are increasingly rare. Recent estimates of the occurrence rate of USP planets are $\sim$0.5$\%$ (e.g. \citealp{sanchis-ojeda2014, winn2018, zhu2021, uzsoy2021}), which is comparable with the frequency of hot Jupiters (e.g. \citealp{howard2012, fressin2013, wang2015, zhu2021}). Additionally, more than 80$\%$ of the known USP population have radii $R_{\mathrm{p}}<2 \, R_{\oplus}$.
There are only eight USP Jupiter-size and/or Jupiter-mass planets with $0.75<P<1 \, \mathrm{d}$. Both the frequency and the small size of USP planets have suggested that they could be the remnant rocky cores of evaporated hot Jupiters (e.g. \citealp{valsecchi2014, konigl2017}). The lack of a metallicity trend for the USP planet host stars, unlike hot Jupiters, suggests that the USP planet population does not coincide with that of hot Jupiters in a later evolutionary stage (e.g. \citealp{fischer2005, winn2017}). Other proposed pathways include photoevaporation of gaseous sub-Neptunes (e.g. \citealp{lundkvist2016, lee2017}), in situ formation \citep{chiang2013}, and inward migration of rocky planets \citep{petrovich2019}. The last hypothesis finds empirical support as several USP planets are part of compact systems showing a broad range of mutual inclinations \citep{petrovich2020}.

In this paper, we report the discovery of two potentially rocky USP planets around M dwarfs from the Transiting Exoplanet Survey Satellite (TESS) and ground-based follow-up programs. During the preparation of this manuscript, \cite{giacalone2022} published another paper  validating TOI-1442 b and TOI-2445 b, among other targets. Here we provide additional evidence  towards the confirmation of both planets through colour contamination analysis of the photometric time series, and measurements of the stellar RVs and cross-correlation functions (CCFs) obtained from new spectroscopic  datasets.

This paper is structured as follows. Section \ref{sec:observations} describes the observations. In particular, Section \ref{sec:obs_tess} describes the TESS observations, Section \ref{sec:obs_ground_phot} the ground-based transit photometry, and Section \ref{sec:spectroscopic_IRD} the spectroscopic observations. Section \ref{sec:star} discusses the characterisation of the host stars. Sections \ref{sec:lc_analysis} and \ref{sec:spectral} explain the methods used to analyse the transit photometric observations and spectroscopic RVs. Section \ref{sec:results} reports the results of our analyses. Section \ref{sec:discussion} puts them into   scientific context. Section \ref{sec:conclusions} summarises our findings.

\begin{figure*}
\centering
\includegraphics[width=0.26\hsize]{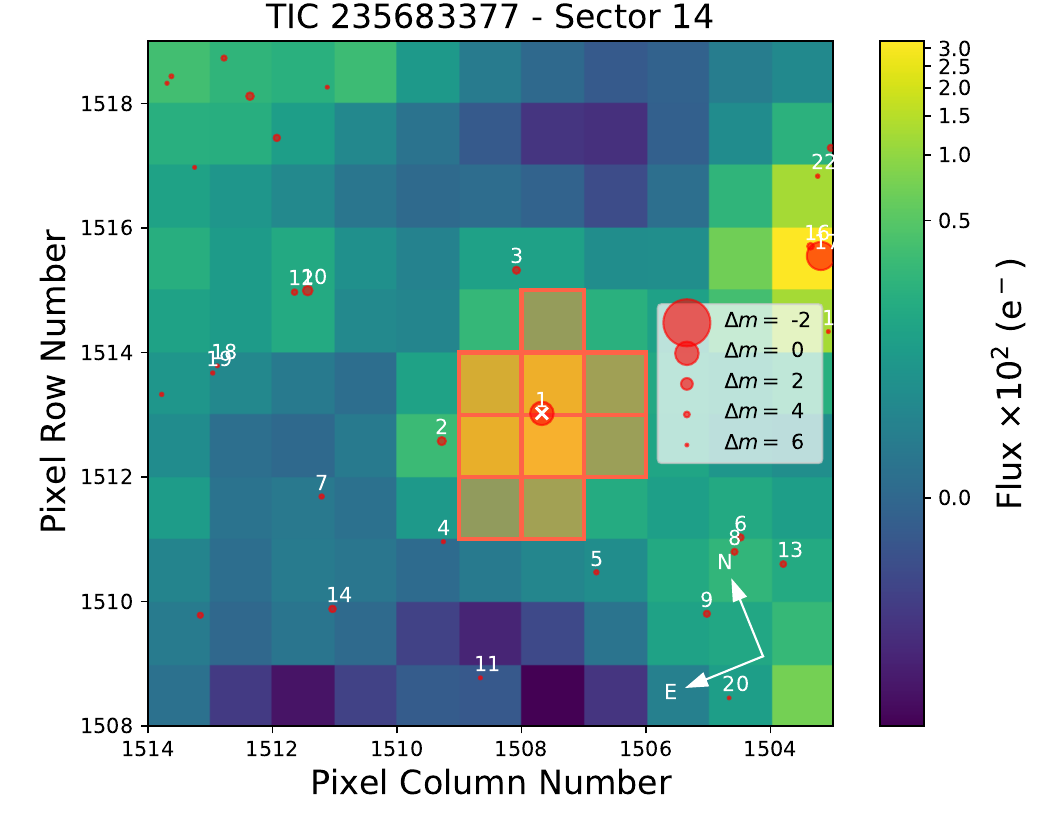}
\includegraphics[width=0.26\hsize]{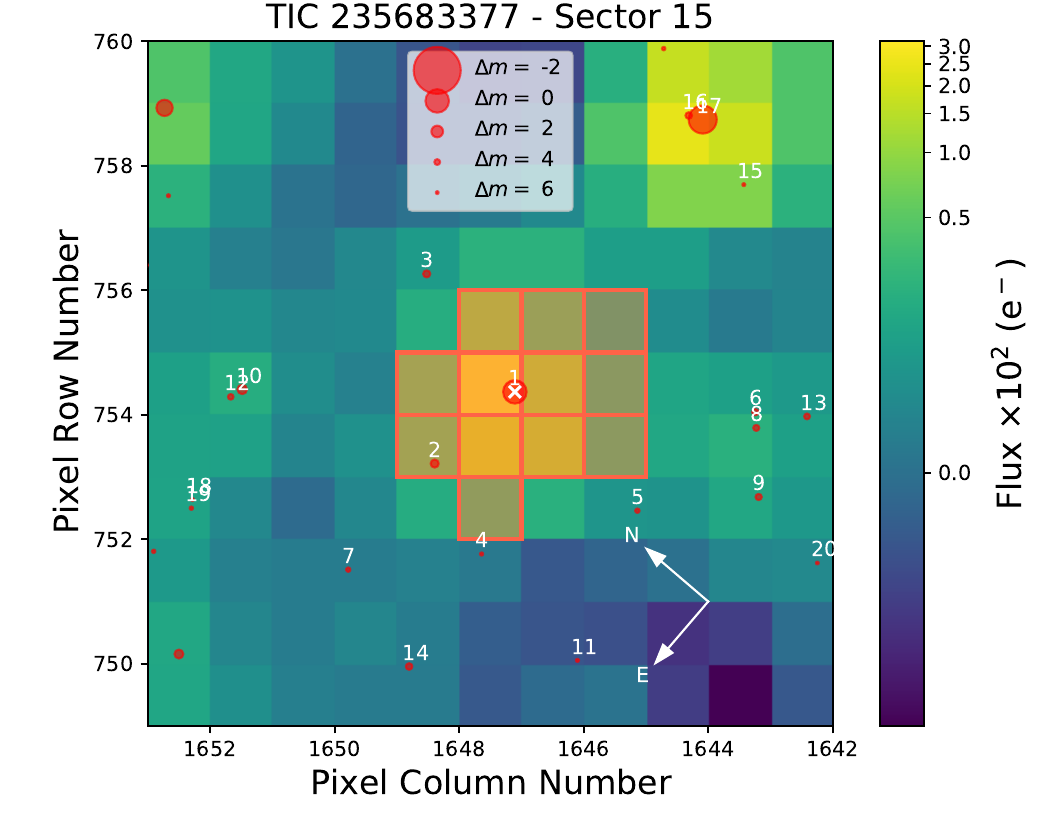}
\includegraphics[width=0.26\hsize]{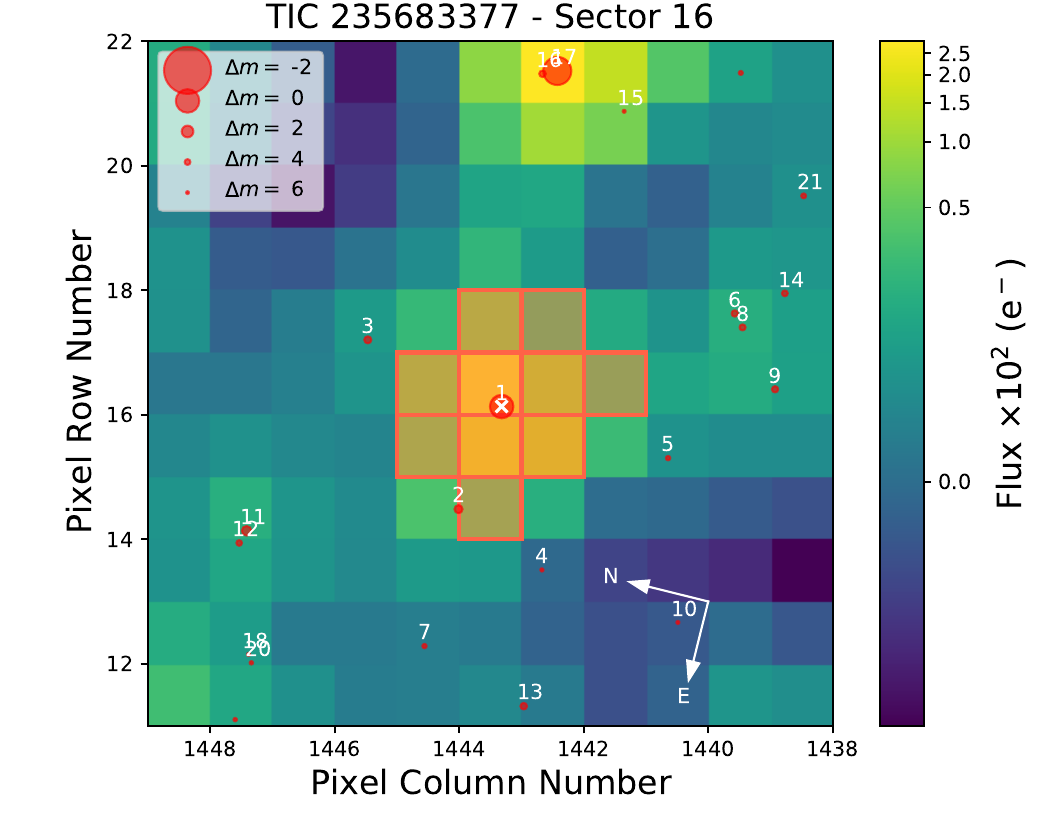}
\includegraphics[width=0.26\hsize]{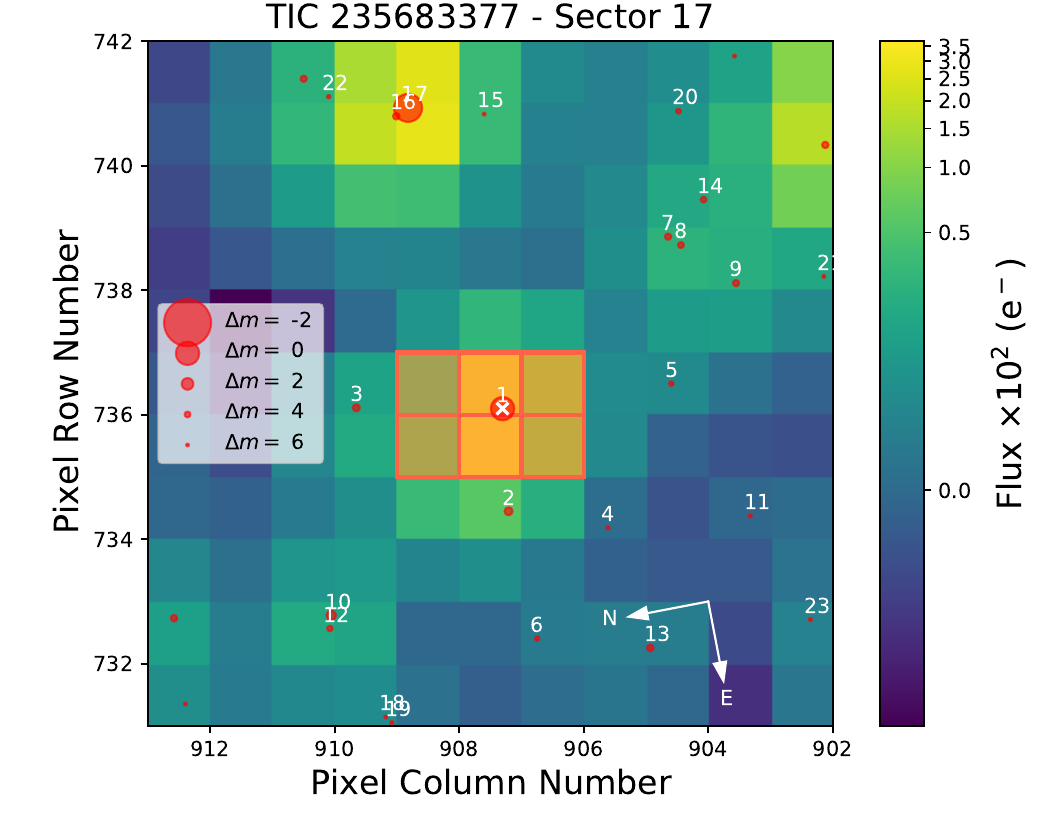}
\includegraphics[width=0.26\hsize]{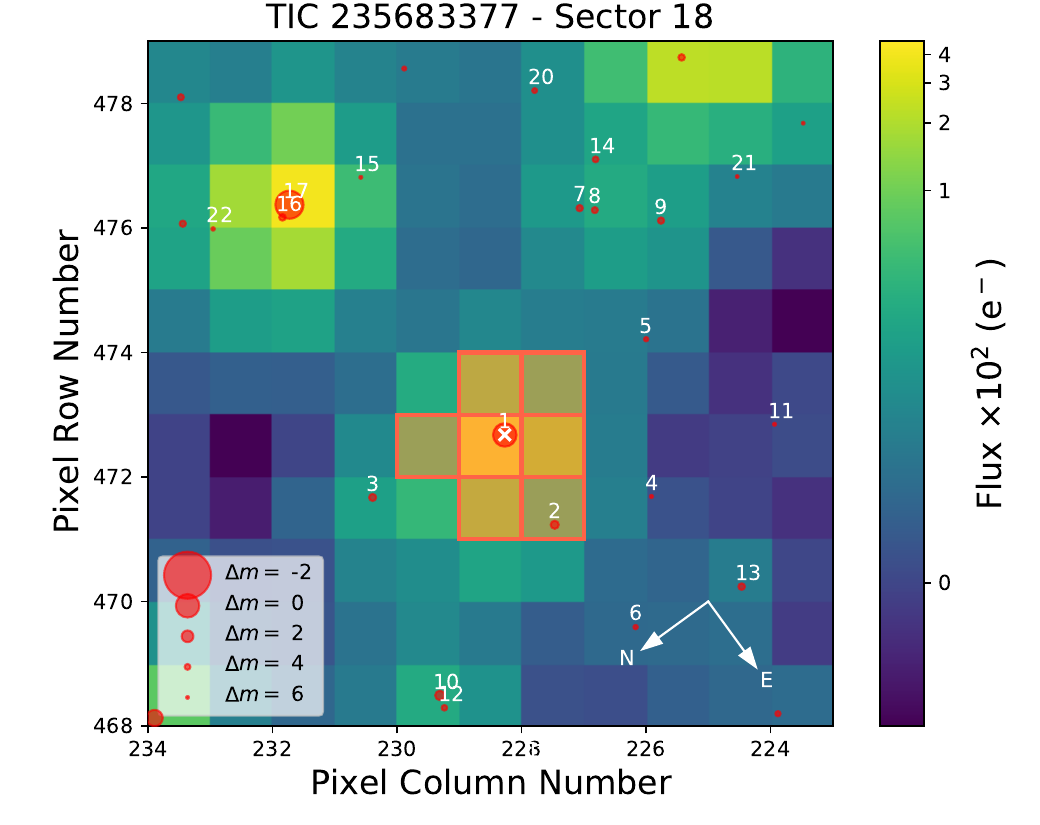}
\includegraphics[width=0.26\hsize]{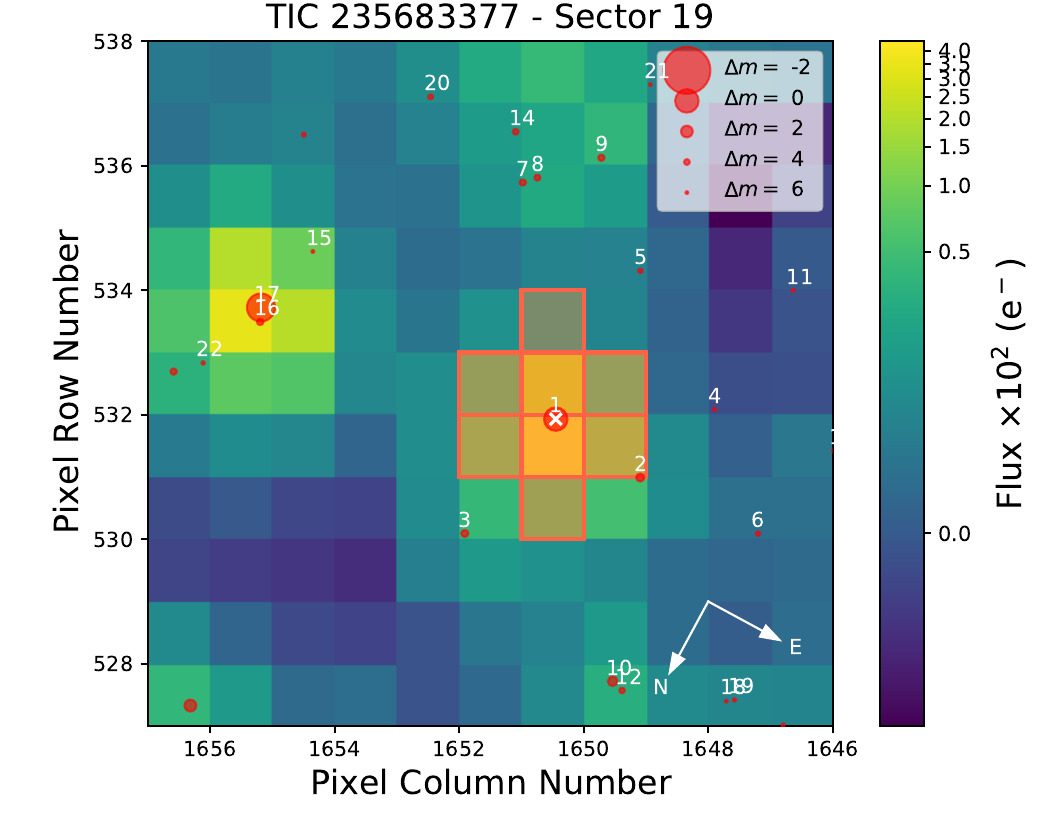}
\includegraphics[width=0.26\hsize]{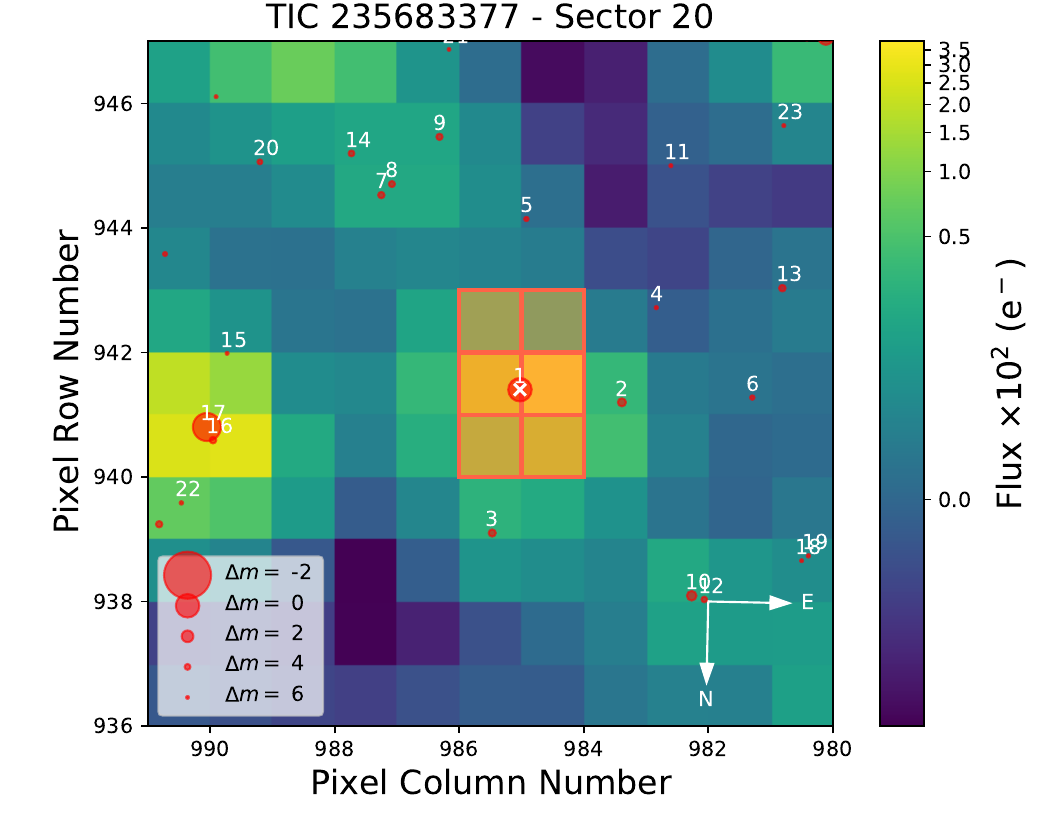}
\includegraphics[width=0.26\hsize]{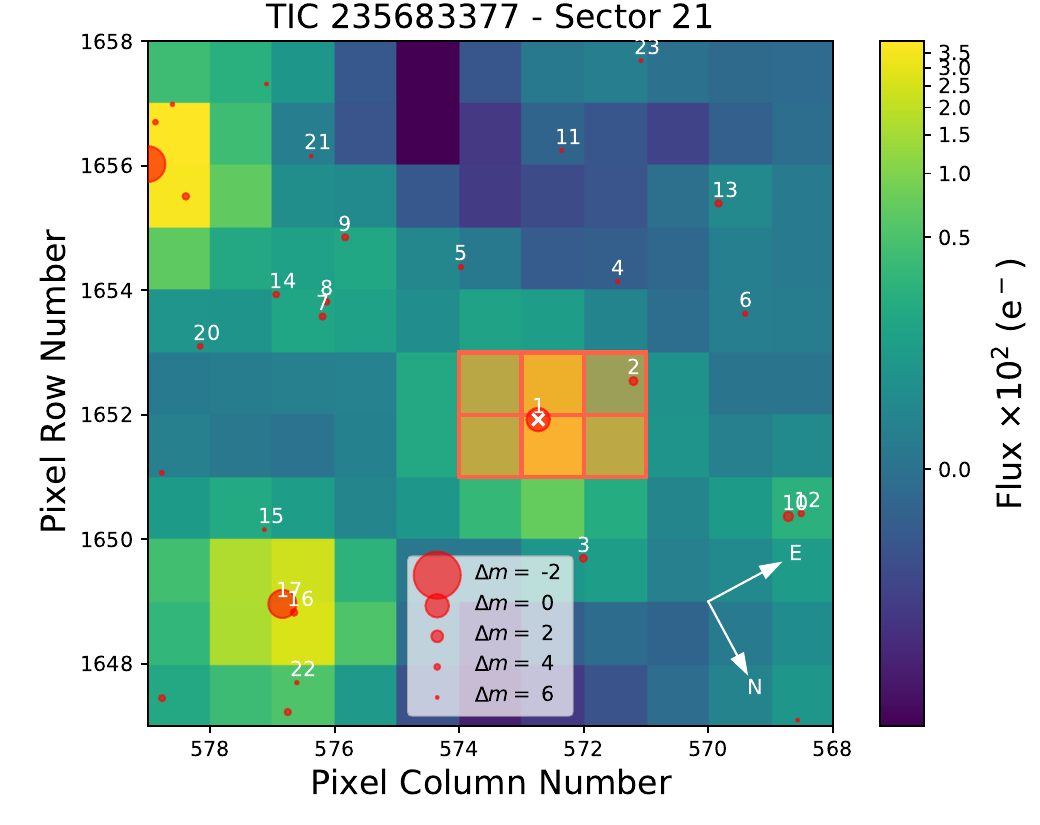}
\includegraphics[width=0.26\hsize]{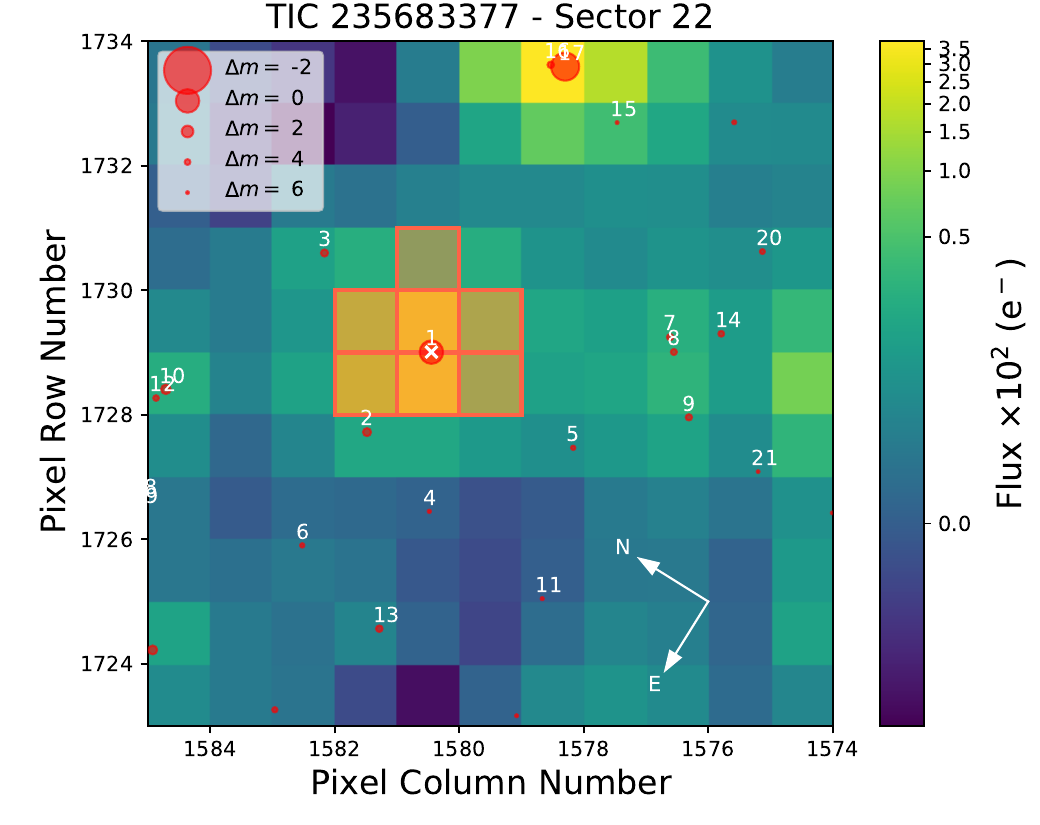}
\includegraphics[width=0.26\hsize]{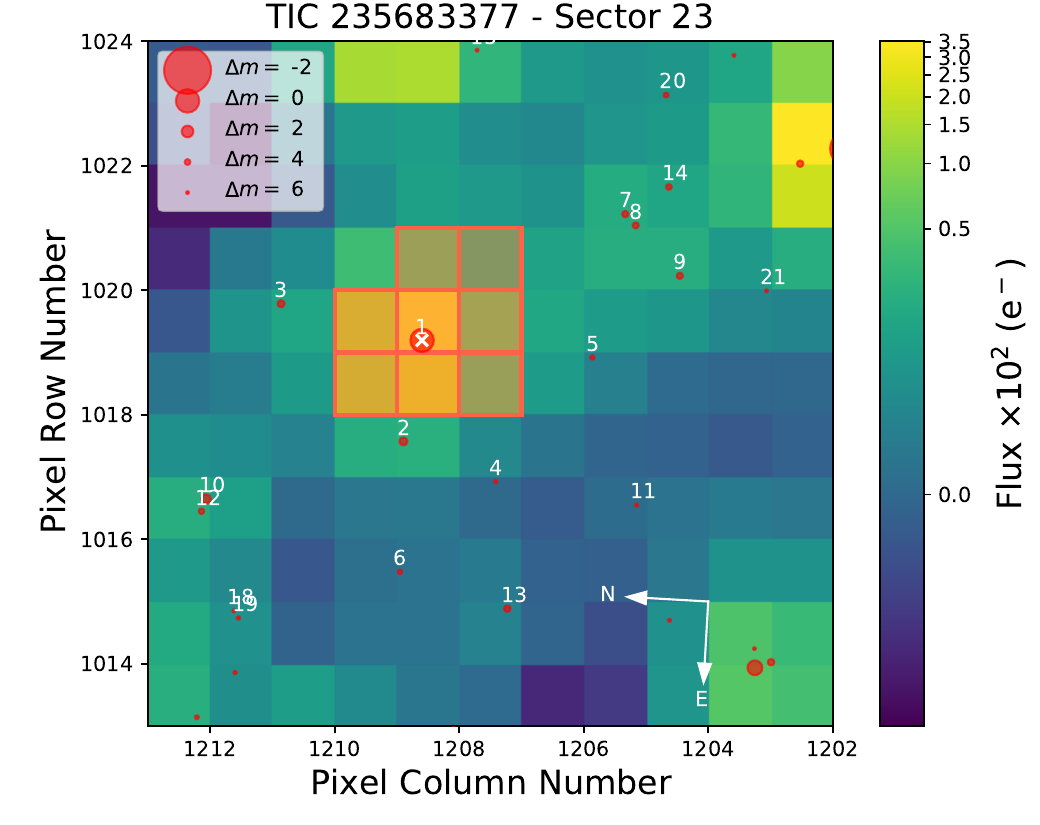}
\includegraphics[width=0.26\hsize]{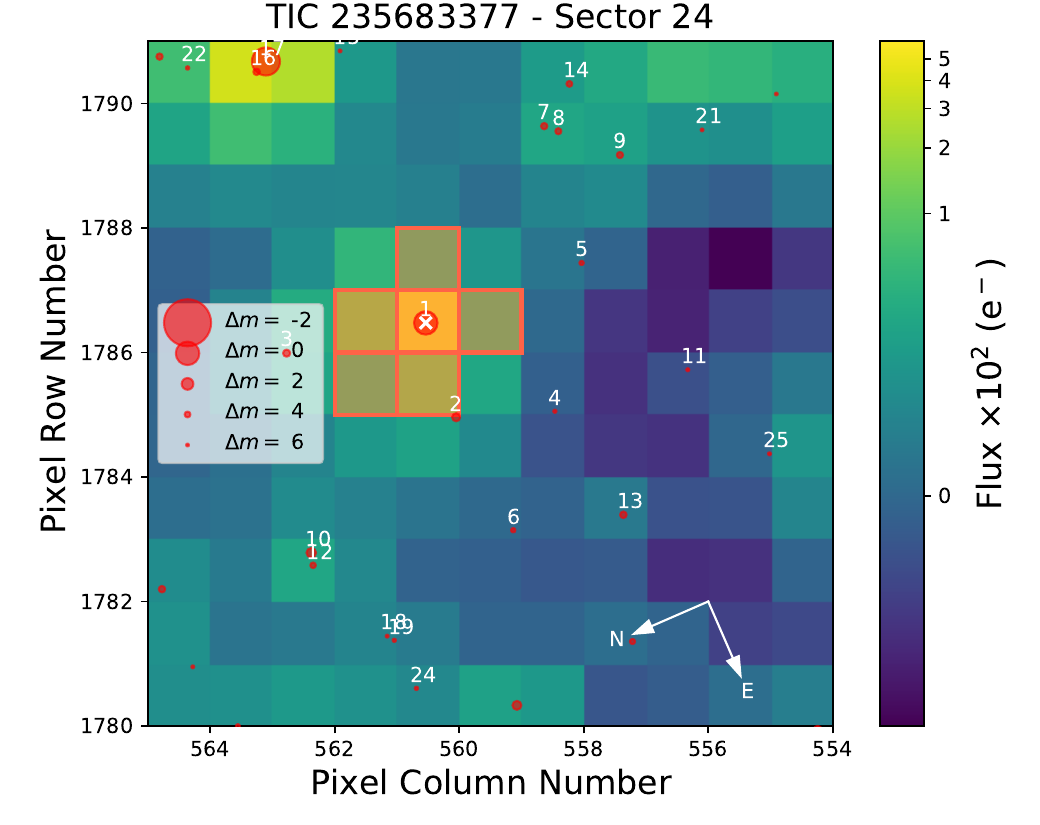}
\includegraphics[width=0.26\hsize]{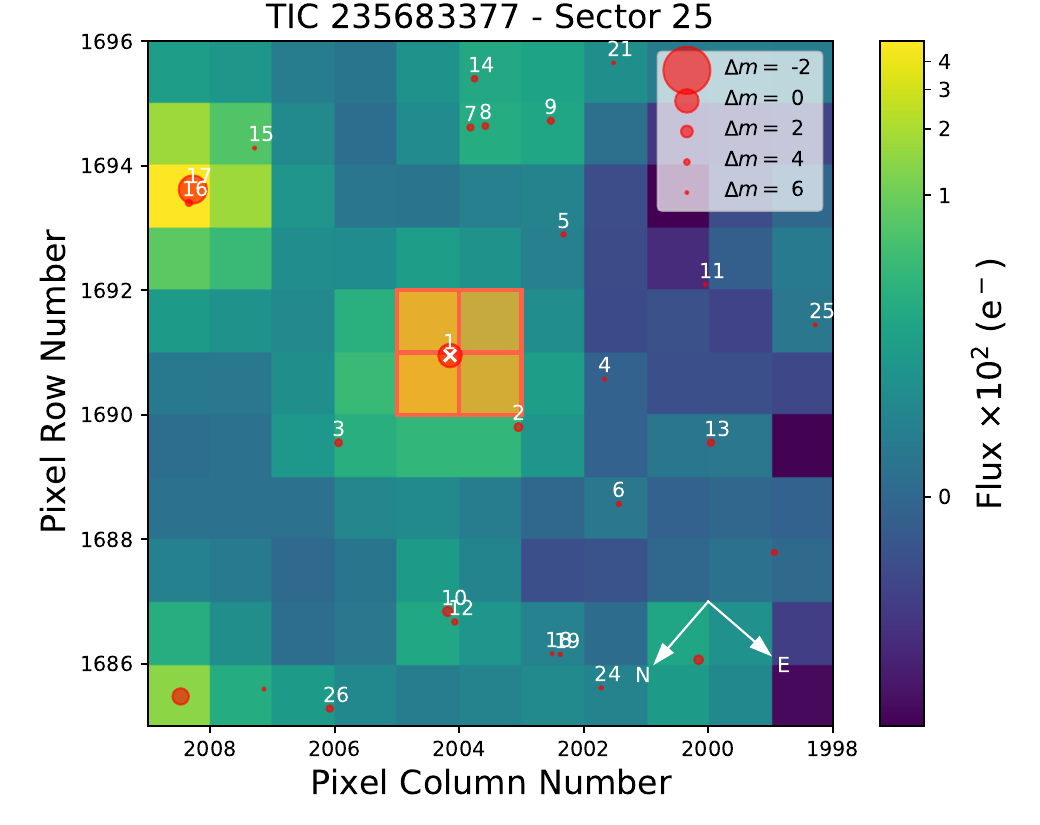}
\includegraphics[width=0.26\hsize]{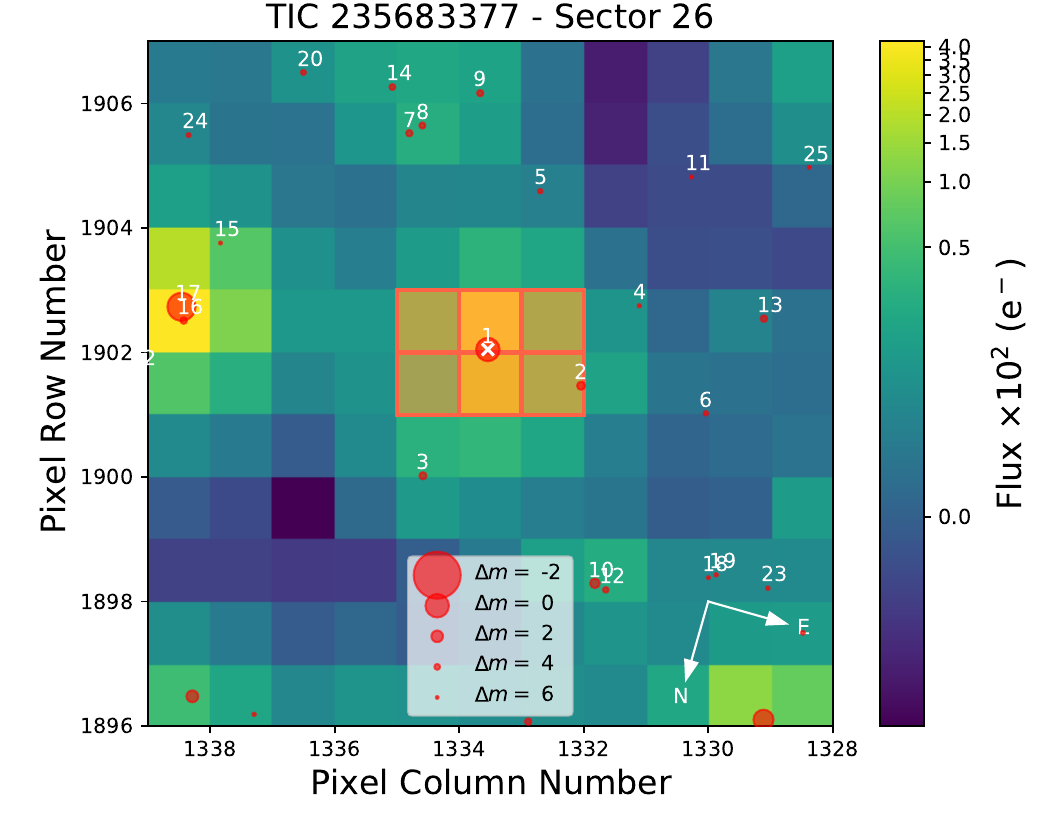}
\includegraphics[width=0.26\hsize]{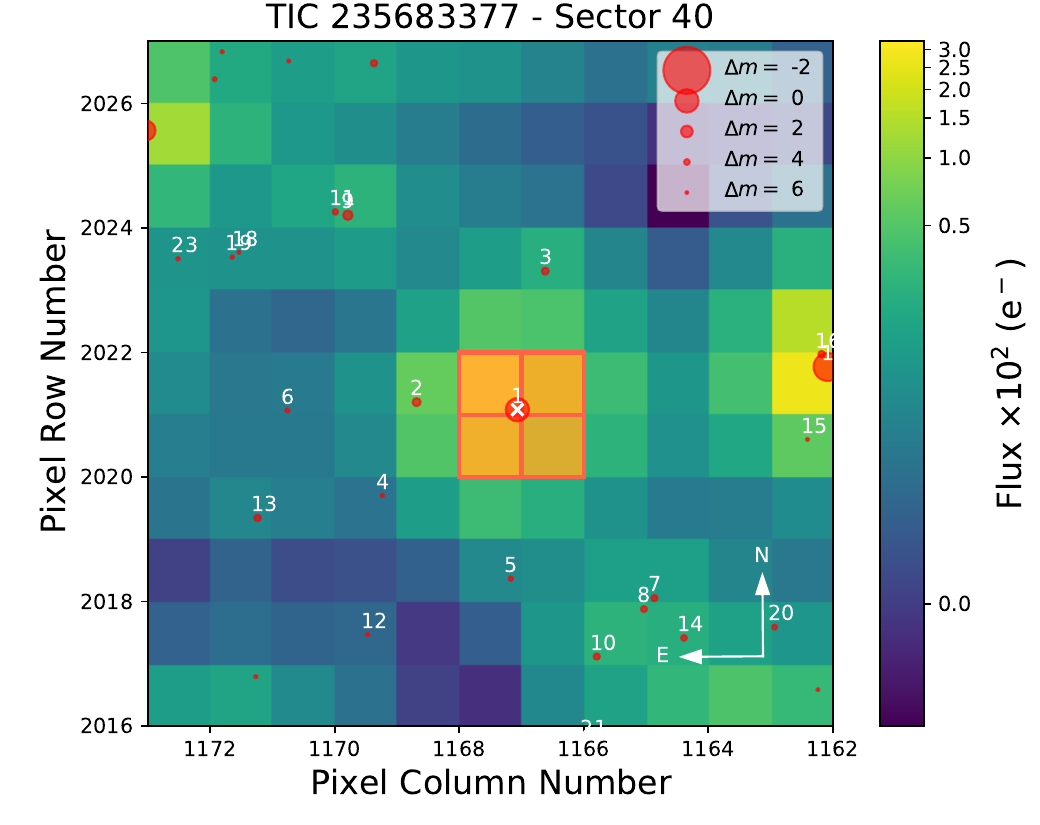}
\includegraphics[width=0.26\hsize]{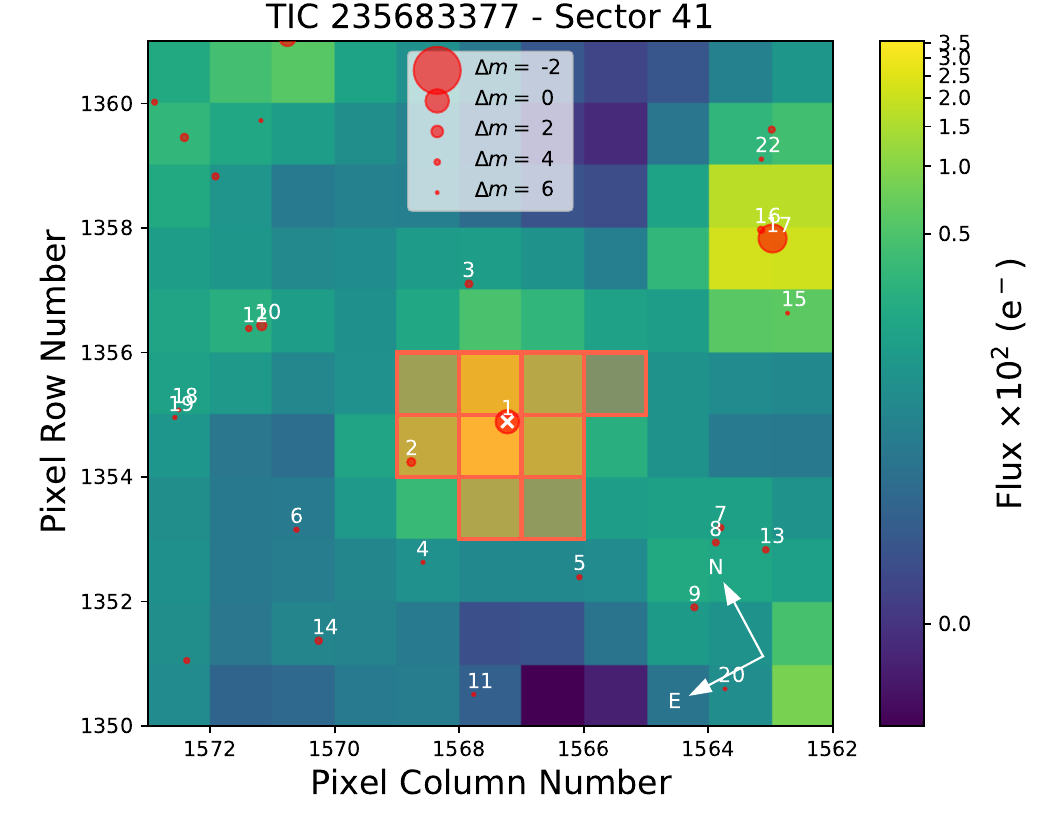}
\includegraphics[width=0.26\hsize]{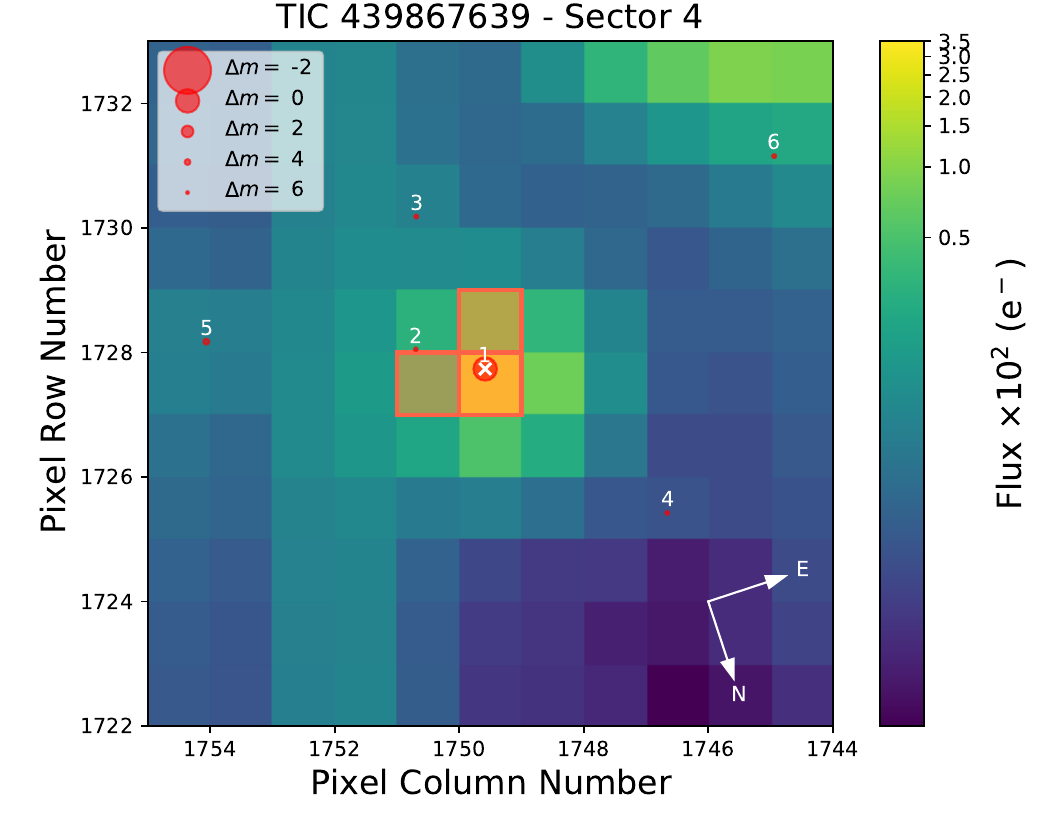}
\includegraphics[width=0.26\hsize]{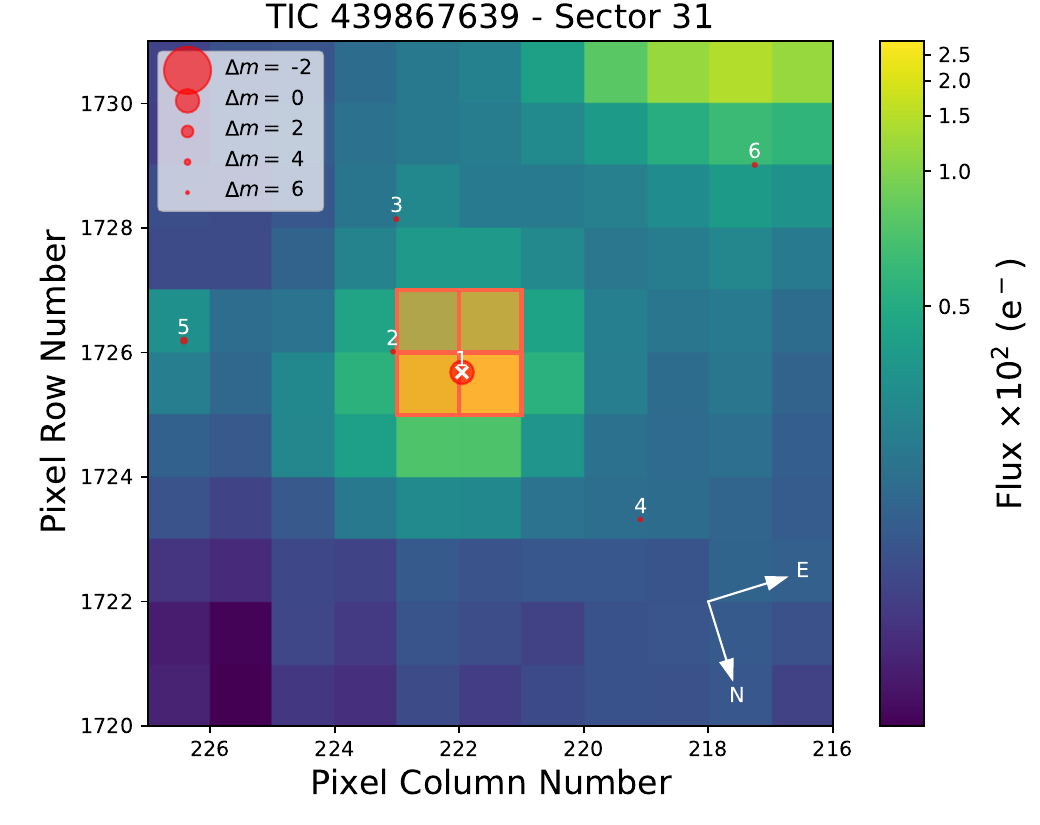}
\includegraphics[width=0.26\hsize]{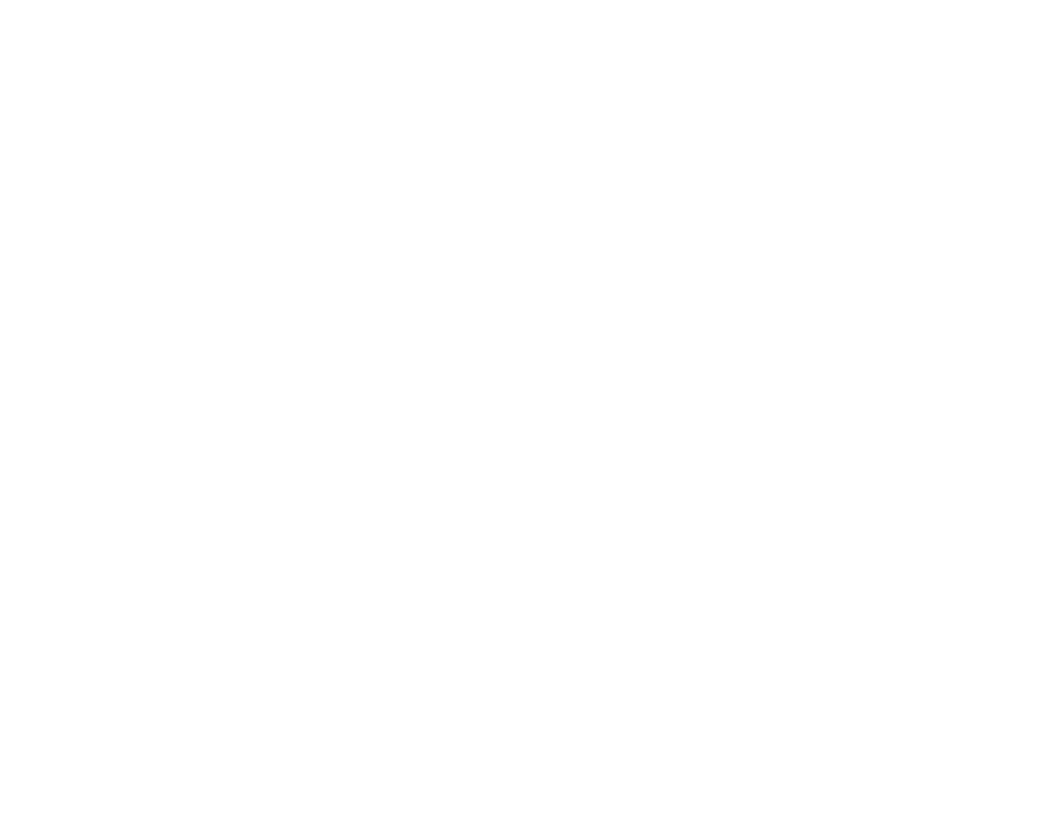}
\caption{TESS target pixel file images of TOI-1442 (top 15 panels) and TOI-2445 (last 2 panels). The pixels highlighted in red denote the aperture mask used to calculate the SAP. The red circles represent neighbouring sources listed in Gaia DR2; the target star is shown with a white `x' and identifier 1. The size of the red circles is inversely proportional to the apparent magnitude difference with respect to the target star. The maximum contrast magnitude of the plotted sources is $\Delta m = 6$ mag.
}
\label{fig:tpf_plots}
\end{figure*}

\section{Observations}
\label{sec:observations}

\begin{table*}[]
\caption{List of ground-based transit observations, their main characteristics, and auxiliary parameters used for detrending.}
\centering
\begin{tabular}{ccccccccc}
\hline
Target & Instrument & UT Date & Passband & Exposure & Dispersion\tablefootmark{a} & Detrending\tablefootmark{b}\\
 & & & & ($s$) & (per cent / exp) & \\
\hline
\noalign{\smallskip}
TOI-1442 & MuSCAT3 & 2021 May 05 & $r$ & 67 & 0.14 & AIRMASS \\
TOI-1442 & MuSCAT3 & 2021 May 05 & $i$ & 30 & 0.20 & AIRMASS \\
TOI-1442 & MuSCAT3 & 2021 May 05 & $z_s$ & 29 & 0.19 & AIRMASS \\
TOI-1442 & MuSCAT3 & 2021 Jun 06 & $r$ & 67 & 0.16 & tot\_C\_cnts \\
TOI-1442 & MuSCAT3 & 2021 Jun 06 & $i$ & 30 & 0.22 & tot\_C\_cnts \\
TOI-1442 & MuSCAT3 & 2021 Jun 06 & $z_s$ & 29 & 0.21 & width\_T1 \\
TOI-1442 & MuSCAT3 & 2021 Jun 17 & $g$ & 240 & 0.14 & AIRMASS \\
TOI-1442 & MuSCAT3 & 2021 Jun 17 & $r$ & 67 & 0.13 & width\_T1 \\
TOI-1442 & MuSCAT3 & 2021 Jun 17 & $i$ & 30 & 0.14 & AIRMASS \\
TOI-1442 & MuSCAT3 & 2021 Jun 17 & $z_s$ & 29 & 0.14 & AIRMASS \\
TOI-1442 & LCO-McD-1m0 & 2020 Aug 14 & $i$ & 90 & 0.14 & AIRMASS \\
TOI-1442 & LCO-McD-1m0 & 2020 Aug 30 & $I$ & 100 & 0.17 & -- \\
TOI-1442 & LCO-McD-1m0 & 2020 Sep 26 & $i$ & 90 & 0.18 & BJD\_TDB, Sky/Pixel\_T1 \\
TOI-1442 & LCO-McD-1m0 & 2020 Oct 21 & $i$ & 90 & 0.13 & AIRMASS \\
TOI-1442 & PESTO & 2020 Feb 09 & $i$ & 30 & 0.24 & AIRMASS \\
\hline
\noalign{\smallskip}
TOI-2445 & MuSCAT & 2021 Feb 07 & $g$ & 30 & 1.44 & BJD\_TDB, AIRMASS \\
TOI-2445 & MuSCAT & 2021 Feb 07 & $r$ & 20 & 0.71 & BJD\_TDB, AIRMASS \\
TOI-2445 & MuSCAT & 2021 Feb 07 & $z_s$ & 35 & 0.15 & BJD\_TDB, AIRMASS \\
TOI-2445 & MuSCAT2 & 2021 Aug 06 & $g$ & 60 & 0.52 & -- \\
TOI-2445 & MuSCAT2 & 2021 Aug 06 & $r$ & 60 & 0.28 & -- \\
TOI-2445 & MuSCAT2 & 2021 Aug 06 & $i$ & 60 & 0.30 & -- \\
TOI-2445 & MuSCAT2 & 2021 Aug 06 & $z_s$ & 60 & 0.23 & -- \\
TOI-2445 & MuSCAT2 & 2021 Sep 14 & $g$ & 70-100 & 0.39 & -- \\
TOI-2445 & MuSCAT2 & 2021 Sep 14 & $i$ & 70 & 0.25 & -- \\
TOI-2445 & MuSCAT2 & 2021 Sep 14 & $z_s$ & 50-60 & 0.25 & -- \\
TOI-2445 & TRAPPIST-S & 2021 Jan 08 & $I+z$ & 50 & 0.42 & -- \\
TOI-2445 & TRAPPIST-S & 2021 Jan 14 & $I+z$ & 50 & 0.37 & -- \\
TOI-2445 & MLO-Lewin-0m36 & 2021 Jan 10 & $I$ & 75 & 1.30 & AIRMASS, Y(FITS)\_T1 \\
\hline
\end{tabular}
\tablefoot{
\tablefoottext{a}{The rms of fitting residuals.}
\tablefoottext{b}{Names of the linear detrending parameters, as given in the data files from ExoFOP-TESS.}
}
\label{tab:detr_aux_params}
\end{table*}

\subsection{TESS photometry}
\label{sec:obs_tess}
TOI-1442 (TIC 235683377) was observed by TESS in 2 min integrations during sectors 14 to 26 (2019 July 18--2020 July 04) and 40 to 41 (2021 June 25--2021 August 20).
It was announced on 2019 November 14 as a TESS object of interest (TOI) by the Science Processing Operations Center (SPOC) pipeline at NASA Ames Research Center.

TOI-2445 (TIC 439867639) was observed by TESS in 2 min integrations during sectors 4 (2018 October 19--2018 November 14) and 31 (2020 October 22--2020 November 16).
It was announced on 2021 January 6 as a TOI also detected  by SPOC.

We downloaded the data from the Mikulski Archive for Space Telescopes\footnote{\url{https://mast.stsci.edu}} (MAST) and extracted the Pre-search Data Conditioned Simple Aperture Photometry (PDCSAP) from the light curve files (extension: `\_lc.fits'). The PDCSAP time series were computed by the SPOC pipeline \citep{jenkins2016spoc}, which calibrates the image data, performs quality control (e.g. flags bad data), calculates the flux for each target in the field of view through simple aperture photometry (SAP module, \citealp{morris2020sap}), and corrects for instrumental systematic effects (PDC module, \citealp{smith2012pdc,stumpe2014pdc}).
Figure \ref{fig:tpf_plots} shows the target pixel files (TPFs) and aperture masks used to extract the SAP flux, created using \texttt{tpfplotter}\footnote{\url{https://github.com/jlillo/tpfplotter}} \citep{aller2020tpfplotter}. There are no known contaminants falling within the aperture masks of either target in most sectors with contrast magnitude $\Delta m < 6$ mag, except for a fainter source with $\Delta m = 3.04$ mag with respect to TOI-1442 in sectors 15, 18, 21, 26, and 41. This contamination was already removed by the PDC pipeline \citep{smith2012pdc,stumpe2014pdc}.

\subsection{Ground-based photometry}
\label{sec:obs_ground_phot}
The TESS Follow-up Observing Program (TFOP) is a network of observatories and researchers whose aims are  to validate the planetary nature of transit-like signals tagged as TOIs, and to measure the masses and radii of the planets,  the orbital properties, and the stellar host parameters. We make use of transit photometry data acquired under the TFOP to confirm that the planetary transits occur on the targeted stars, to rule out some false positive scenarios (e.g. blended eclipsing binaries), and to refine the planet's radius measurement. The TFOP data are available to the working group members on the Exoplanet Follow-up Observing Program for TESS (ExoFOP-TESS)\footnote{\url{https://exofop.ipac.caltech.edu/tess/}} website. In Table~\ref{tab:detr_aux_params} we summarise the ground-based photometric observations analysed in this paper.

\subsubsection{MuSCAT}
The Multi-colour Simultaneous Camera for studying Atmospheres of Transiting planets (MuSCAT, \citealp{narita2015muscat}) is a three-band imager mounted on the 188 cm telescope of National Astronomical Observatory of Japan in Okayama, Japan. MuSCAT is equipped with three 1024$\times$1024 pixel CCDs with a pixel scale of 0$''$.358 pixel$^{-1}$. The field of view of MuSCAT is 6$'$.1$\times$6$'$.1. Each CCD is coupled with an Astrodon Photometrics Generation 2 Sloan filter. The three filter bands are $g$ (400--550 nm), $r$ (550--700 nm) and $z_s$ (820--920 nm).

We observed a full transit of TOI-2445 b on 2021 February 7 UT with MuSCAT. The exposure times were 30, 20, and 35 s for the $g$, $r$,  and $z_s$ bands, respectively. After performing dark and flat-field calibrations, we extracted light curves by aperture photometry with radii of 3$''$.6, 4$''$.3, and 3$''$.6 for the respective filters using a custom data reduction pipeline \citep{fukui2011}. The resulting photometric dispersion per exposure (i.e. the root mean square   of the light curve fitting residuals) are 1.44$\%$, 0.71$\%$, and 0.15$\%$. 

\subsubsection{MuSCAT2}
MuSCAT2 \citep{narita2019muscat2} is a four-band imager mounted on the 152 cm Telescopio Carlo S{\'a}nchez (TCS) at Teide Observatory in Spain. MuSCAT2 is equipped with four 1024$\times$1024 pixel CCDs with a pixel scale of 0$''$.435 pixel$^{-1}$. The field of view of MuSCAT2 is 7$'$.4$\times$7$'$.4. The four filter bands are $g$ (400--550 nm), $r$ (550--700 nm), $i$ (700--820 nm), and $z_s$ (820--920 nm). The $g$, $r$, and $z_s$ filters are identical to those adopted by MuSCAT. The $i$ filter was custom-ordered and manufactured by Asahi Spectra Co., Ltd.

We observed a partial transit of TOI-2445 b on 2021 August 6 UT and a full transit on 2021 September 14 UT with MuSCAT2. The $r$ filter was not available during the last observation. The exposure times were set between 50 and 100 s, depending on the band, instrument, and observing conditions. We extracted photometry with aperture radii of 10$''$.875. The resulting photometric dispersion per exposure are 0.52-0.39$\%$ in $g$, 0.28$\%$ in $r$, 0.30-0.25$\%$ in $i$, and 0.23-0.25$\%$ in $z_s$ for the two nights, respectively.

\subsubsection{MuSCAT3}
MuSCAT3 \citep{narita2020muscat3} is another four-band imager with $g$, $r$, $i$, and $z_s$ filters, mounted on the 2 m Faulkes Telescope North (FTN) of Las Cumbres Observatory at the Haleakala observatory in Hawaii. The 2048$\times$2048 pixel CCDs, with a pixel scale of 0$''$.266 pixel$^{-1}$, were manufactured by Teledyne Princeton Instruments. The field of view of MuSCAT3 is 9$'$.1$\times$9$'$.1.

We observed three transits of TOI-1442 b on 2021 May 05, 2021 June 06, and 2021 June 17 UT with MuSCAT3. We discarded the $g$-band data from the first two observations because of too high scatter. The exposure times were set to 240, 67, 30, and 29 s for the $g$, $r$, $i$, and $z_s$ bands, respectively. We extracted photometry with aperture radii of 2$''$.65, 7$''$.95, and 4$''$.68 for the respective filters. The resulting photometric dispersion per exposure are 0.14$\%$ in $g$, 0.13-0.16$\%$ in $r$, 0.14-0.22$\%$ in $i$, and 0.14-0.21$\%$ in $z_s$.

\subsubsection{TRAPPIST-South}
The TRAnsiting Planets and PlanetesImals Small Telescope-South (TRAPPIST-South, \citealp{jehin2011trappist,Gillon2011}) is a 60 cm telescope at La Silla Observatory in Chile. It is equipped with a  thermoelectrically cooled $2K\times2K$ pixel FLI Proline PL3041-BB CCD with a pixel scale of 0$''$.64 pixel$^{-1}$, resulting in a field of view of  22$'\times$22$'$.

We observed two transits of TOI-2445\,b on 2021 January 8 and 14 in the $I+z$ filter with an exposure time of 50~s. We used the {\tt TESS Transit Finder} tool,\footnote{\url{https://astro.swarthmore.edu/transits/}} which is a customised version of the {\tt Tapir} software package \citep{Jensen_2013ascl}, to schedule the photometric  time series. Data calibration and photometric measurements were performed using the \textit{PROSE}\footnote{\url{https://github.com/lgrcia/prose}} pipeline \citep{garcia2021}. The resulting photometric dispersions per exposure are 0.42--0.37$\%$ for the two nights.

\subsubsection{LCOGT 1\,m}

We observed four full transits of TOI-1442 b from the Las Cumbres Observatory Global Telescope \citep[LCOGT;][]{Brown:2013} 1.0\,m network node at McDonald Observatory. Observations on 2020 August 14 and 2020 August 30 were conducted in I band  with exposure times of 100$\,s$, and observations on 2020 September 26 and 2020 October 21 were conducted in Sloan $i'$ band with exposure times of 90$\,s$. We used the {\tt TESS Transit Finder} to schedule our transit observations. The 1\,m telescopes are equipped with $4096\times4096$ pixel SINISTRO cameras with an image scale of $0\farcs389$ per pixel, resulting in a $26\arcmin\times26\arcmin$ field of view. The images were calibrated by the standard LCOGT {\tt BANZAI} pipeline \citep{McCully:2018}, and photometric data were extracted using {\tt AstroImageJ} \citep{Collins:2017}. The images were focused and have mean stellar point spread functions with a FWHM of $\sim2\arcsec$, and circular photometric apertures with radius $\sim4\arcsec$ were used to extract the differential photometry. The apertures exclude flux from the nearest Gaia EDR3 and TESS Input Catalog neighbour (TIC 1718221659) $13\arcsec$ west of the target. The resulting photometric dispersions per exposure are in the range 0.13--0.18$\%$.

\subsubsection{OMM/PESTO}

We observed a full transit of TOI-1442 b at Observatoire du Mont-M\'{e}gantic, Canada, on 2020 February 9. The observations were made in the $i^\prime$ filter using the 1.6 m telescope of the observatory equipped with the PESTO camera. The adopted exposure times were of 30$\,s$. The light curve extraction via differential photometry was accomplished with {\tt AstroImageJ}, which was also used for image calibration (bias subtraction and flatfield division). The images have typical stellar point spread functions with a FWHM of $\sim2\arcsec$, and circular photometric apertures with radius $\sim5\arcsec$ were used to extract the differential photometry, excluding flux from the nearest Gaia EDR3 neighbour. The resulting photometric dispersion per exposure is 0.24$\%$.

\subsubsection{MLO/Lewin}

We observed a full transit of TOI-2445 b in I band on 2021 January 10 from the Maury Lewin Astronomical Observatory 0.36\,m telescope near Glendora, CA. The telescope is equipped with $3326\times2504$ pixel SBIG STF8300M camera having an image scale of $0\farcs84$ per pixel, resulting in a $23\arcmin\times17\arcmin$ field of view. The adopted exposure times were of 75$\,s$. The images were calibrated and the photometric data were extracted using {\tt AstroImageJ}. The images have typical stellar point spread functions with a FWHM of $\sim5\arcsec$, and circular photometric apertures with radius $\sim7\arcsec$ were used to extract the differential photometry, which excluded most of the flux from the nearest Gaia EDR3 neighbour. The resulting photometric dispersion per exposure is 1.30$\%$.

\subsection{Spectroscopic observations with Subaru/IRD}
\label{sec:spectroscopic_IRD}

The InfraRed Doppler (IRD) instrument on the 8.2m Subaru telescope is a fibre-fed spectrograph covering the wavelength range 930--1740 nm with a spectral resolution of $R\sim$70\,000 \citep{tamura2012,kotani2018}.
We obtained 20 IRD spectra of TOI-1442 on 11 nights between 2020 September 27 and 2021 October 22 UT. We also obtained 12 IRD spectra of TOI-2445 on five nights between 2021 September 9 and 2021 November 12 UT.
The exposure times were set to 1200--1800 s for TOI-1442 and 900--1800 s for TOI-2445, depending on the observing conditions. For all scientific exposures, we also injected the laser-frequency comb into the spectrograph, whose spectra were used for the simultaneous wavelength calibrations. 

We reduced raw IRD data and extracted one-dimensional stellar and reference (laser-frequency comb) spectra as described in \citet{2020PASJ...72...93H}. Wavelengths were calibrated by using the data of the Thorium-Argon hollow-cathode lamp as well as the laser-frequency comb taken during the daytime. 
Using IRD's standard analysis pipeline \citep{2020PASJ...72...93H}, we measured RVs for both TOI-1442 and TOI-2445. In short, we extracted a template spectrum for each target that is free from telluric features and instrumental broadening based on multiple observed frames, and relative RVs were measured with respect to this template by the forward modelling technique taking into account the instantaneous instrumental profile of the spectrograph and telluric absorptions. 
The resulting RV precisions (internal errors) were typically $4-5$ m s$^{-1}$ for TOI-1442 and $5-8$ m s$^{-1}$ for TOI-2445.

\begin{table*}[]
\caption{Stellar properties of TOI-1442.}
\centering
\begin{tabular}{ccccc}
\hline
Parameter & \multicolumn{3}{c}{Value} & Reference  \tablefootmark{a}\\
\hline
\multicolumn{5}{c}{Astrometric Parameters} \\
\noalign{\smallskip}
$\alpha$ (epoch 2016.0) & \multicolumn{3}{c}{19:9:10.1142} & (1)\\
\noalign{\smallskip}
$\delta$ (epoch 2016.0) & \multicolumn{3}{c}{+74:10:27.626} & (1)\\
\noalign{\smallskip}
$\mu_\alpha$ (mas/yr) & \multicolumn{3}{c}{$81.959 \pm 0.020$} & (1)\\
\noalign{\smallskip}
$\mu_\delta$ (mas/yr) & \multicolumn{3}{c}{$462.708 \pm 0.017$} & (1)\\
\noalign{\smallskip}
Parallax (mas) & \multicolumn{3}{c}{$24.164 \pm 0.015$} & (1)\\
\noalign{\smallskip}
Distance (pc) & \multicolumn{3}{c}{$41.316 \pm 0.023$} & (2)\\
\hline
\multicolumn{5}{c}{Photospheric Parameters} \\
\noalign{\smallskip}
 & Spec. Synth. & SED fit & Emp. Rel. & \\
 \noalign{\smallskip}
$T_{\mathrm{*,eff}}$ ($K$) \tablefootmark{b} & 3345$\pm$100 & 3263$\pm$100 & & This work \\
\noalign{\smallskip}
$\log{g_*}$ (dex) & & 5.27$_{-0.37}^{+0.17}$ & 4.91$\pm$0.03 & This work \\
\noalign{\smallskip}
$[M/H]_*$ (dex) & 0.10$\pm$0.07 & 0.12$\pm$0.07 & & This work \\
\noalign{\smallskip}
$[Fe/H]_*$ (dex) & 0.03$\pm$0.15 & & & This work \\
\hline
\multicolumn{5}{c}{Physical Parameters} \\
\noalign{\smallskip}
 & Spec. Synth. & SED fit & Emp. Rel. & \\
 \noalign{\smallskip}
$M_*$ ($M_{\odot}$) & & & 0.2843$\pm$0.0065 & This work \\
\noalign{\smallskip}
$R_*$ ($R_{\odot}$) & & 0.3159$_{-0.0047}^{+0.0036}$ & 0.3105$\pm$0.0089 & This work \\
\noalign{\smallskip}
$L_*$ ($L_{\odot}$) & & 0.01020$_{-0.00012}^{+0.00014}$ & & This work \\
\noalign{\smallskip}
$\rho_*$ ($\rho_{\odot}$) & & & 9.5$\pm$0.8 & This work \\
\hline
\multicolumn{5}{c}{Magnitudes} \\
\noalign{\smallskip}
$V$ (mag) & \multicolumn{3}{c}{15.39$\pm$0.20} & (3)\\
\noalign{\smallskip}
$G$ (mag) & \multicolumn{3}{c}{13.7390$\pm$0.0029} & (1)\\
\noalign{\smallskip}
$TESS$ (mag) & \multicolumn{3}{c}{12.4934$\pm$0.0075} & (3)\\
\noalign{\smallskip}
$J$ (mag) & \multicolumn{3}{c}{10.925$\pm$0.019} & (4)\\
\noalign{\smallskip}
$H$ (mag) & \multicolumn{3}{c}{10.332$\pm$0.019} & (4)\\
\noalign{\smallskip}
$K_s$ (mag) & \multicolumn{3}{c}{10.089$\pm$0.020} & (4)\\
\hline
\end{tabular}
\tablefoot{
\tablefoottext{a}{References: (1) Gaia EDR3 \citep{2021A&A...649A...1G}, (2) \citet{bailer-jones2021}, (3) TIC v8.2 \citep{Stassun_2019}, (4) 2MASS \citep{2006AJ....131.1163S}.}\\
\tablefoottext{b}{A systematic error of $\sim$100 $K$ is associated with the absolute calibration of the instruments, as in \cite{kawauchi2022}. This term is dominant compared to the random error from the fitting procedure}.}
\label{tab:toi1442_params}
\end{table*}

\begin{table*}[]
\caption{Stellar properties of TOI-2445.}
\centering
\begin{tabular}{ccccc}
\hline
Parameter & \multicolumn{3}{c}{Value} & Reference \tablefootmark{a}\\
\hline
\multicolumn{5}{c}{Astrometric Parameters} \\
\noalign{\smallskip}
$\alpha$ (epoch 2016.0) & \multicolumn{3}{c}{02:53:15.8792} & (1)\\
\noalign{\smallskip}
$\delta$ (epoch 2016.0) & \multicolumn{3}{c}{+00:03:08.400} & (1)\\
\noalign{\smallskip}
$\mu_\alpha$ (mas/yr) & \multicolumn{3}{c}{$57.557 \pm 0.029$} & (1)\\
\noalign{\smallskip}
$\mu_\delta$ (mas/yr) & \multicolumn{3}{c}{$-23.693
 \pm 0.025$} & (1)\\
 \noalign{\smallskip}
Parallax (mas) & \multicolumn{3}{c}{$20.361 \pm 0.032$} & (1)\\
\noalign{\smallskip}
Distance (pc) & \multicolumn{3}{c}{$48.968 \pm 0.077$} & (2)\\
\hline
\multicolumn{5}{c}{Photospheric Parameters} \\
\noalign{\smallskip}
& Spec. Synth. & SED fit & Emp. Rel. & \\
\noalign{\smallskip}
$T_{\mathrm{*,eff}}$ ($K$) \tablefootmark{b} & 3375$\pm$100 & 3238$\pm$100 & & This work \\
\noalign{\smallskip}
$\log{g_*}$ (dex) & & 5.24$_{-0.40}^{+0.20}$ & 4.96$\pm$0.03 & This work \\
\noalign{\smallskip}
$[M/H]_*$ (dex) & -0.32$\pm$0.07 & -0.34$\pm$0.07 & & This work \\
\noalign{\smallskip}
$[Fe/H]_*$ (dex) & -0.39$\pm$0.16 & & & This work \\
\hline
\multicolumn{5}{c}{Physical Parameters} \\
\noalign{\smallskip}
& Spec. Synth. & SED fit & Emp. Rel. & \\
\noalign{\smallskip}
$M_*$ ($M_{\odot}$) & & & 0.2448$\pm$0.0056 & This work \\
\noalign{\smallskip}
$R_*$ ($R_{\odot}$) & & 0.2762$_{-0.0051}^{+0.0055}$ & 0.2699$\pm$0.0078 & This work \\
\noalign{\smallskip}
$L_*$ ($L_{\odot}$) & & 0.00760$_{-0.00017}^{+0.00019}$ & & This work \\
\noalign{\smallskip}
$\rho_*$ ($\rho_{\odot}$) & & & 12.5$\pm$1.1 & This work \\
\hline
\multicolumn{5}{c}{Magnitudes} \\
\noalign{\smallskip}
$V$ (mag) & \multicolumn{3}{c}{15.69$\pm$0.03} & (3)\\
\noalign{\smallskip}
$G$ (mag) & \multicolumn{3}{c}{14.3977$\pm$0.0006} & (1)\\
\noalign{\smallskip}
$TESS$ (mag) & \multicolumn{3}{c}{13.1292$\pm$0.0076} & (3)\\
\noalign{\smallskip}
$J$ (mag) & \multicolumn{3}{c}{11.555$\pm$0.023} & (4)\\
\noalign{\smallskip}
$H$ (mag) & \multicolumn{3}{c}{11.033$\pm$0.022} & (4)\\
\noalign{\smallskip}
$K_s$ (mag) & \multicolumn{3}{c}{10.779$\pm$0.019} & (4)\\
\hline
\end{tabular}
\tablefoot{
\tablefoottext{a}{References: (1) Gaia EDR3 \citep{2021A&A...649A...1G}, (2) \citet{bailer-jones2021}, (3) TIC v8.2 \citep{Stassun_2019}, (4) 2MASS \citep{2006AJ....131.1163S}.}\\
\tablefoottext{b}{A systematic error of $\sim$100 $K$ is associated with the absolute calibration of the instruments, as in \cite{kawauchi2022}. This term is dominant compared to the random error from the fitting procedure}.}
\label{tab:toi2445_params}
\end{table*}

\section{Stellar characterisation}
\label{sec:star}
We applied the following procedure to determine the main stellar parameters relevant to the light curve analysis. Firstly, we derived the effective temperature ($T_{\mathrm{*,eff}}$), iron abundance ($[Fe/H]_*$), and overall metallicity ($[M/H]_*$) from the telluric-free template IRD spectra by following the same analysis as in \citet{2022AJ....163...72I}.   We note that for TOI-2445, we risked   using the spectrum without instrumental-profile deconvolution. This is because the signal-to-noise ratio (S/N) per frame of TOI-2445 is so low that the noise is amplified during the deconvolution, especially in the $Y$ band.

We used the 47 FeH molecular lines in the Wing-Ford band at $990-1020$ nm to estimate $T_{\mathrm{*,eff}}$. 
We note that the errors given in Tables \ref{tab:toi1442_params} and \ref{tab:toi2445_params} are dominated by systematic errors, which are much larger than the standard deviation ($\sigma$) of estimates based on individual FeH lines divided by the square root of the number of lines ($\sigma / \sqrt{N}$).
More details can be found in \citet{2022AJ....163...72I}.
For elemental abundances, we used 34 and 30 neutral atomic lines for TOI-1442 and TOI-2445, respectively.
The atomic species responsible for these lines are Na, Mg, Ca, Ti, Cr, Mn, Fe, and Sr.
The analyses are based on the equivalent width comparison of individual absorption lines between the calculated model spectra and the observed spectra.
The detailed procedures to determine individual elemental abundances and their errors are described in \citet{2020PASJ...72..102I}.

We iterated the $T_{\mathrm{*,eff}}$ estimation and the abundance analysis  as described below.
Firstly, we derived a provisional $T_{\mathrm{*,eff}}$ adopting the solar metallicity ($[Fe/H]_* = 0$), and then we determined the individual abundances of the eight elements $[X/H]_*$ using the $T_{\mathrm{*,eff}}$. Secondly, we redetermined the  $T_{\mathrm{*,eff}}$ adopting the iron abundance $[Fe/H]_*$ as the input metallicity, and then we redetermined the abundances using the new $T_{\mathrm{*,eff}}$. 
We iterated the estimation of $T_{\mathrm{*,eff}}$ and $[Fe/H]_*$ until the final results and the results of the previous step agreed within the error margin.
From the final results of the abundances of the eight elements, $[M/H]_*$ was determined by calculating the average weighted by the inverse of the square of their estimated errors.

We performed spectral energy distribution (SED) fitting on the photometric magnitudes of Gaia EDR3 $G$, $BP$, and $RP$ bands \citep{2021A&A...649A...1G}; 2MASS $J$, $H$, and $K_s$ bands \citep{2006AJ....131.1163S}; and WISE $W1$, $W2$, and $W3$ bands \citep{2014yCat.2328....0C}, applying the $[M/H]_*$ prior from the spectral analysis. Hence, we obtained an independent estimate of $T_{\mathrm{*,eff}}$, along with the stellar radius ($R_*$) and luminosity ($L_*$).
The BT-Settl synthetic spectra \citep{allard2014} were then fitted to the SED with the parameters of $T_{\rm {*,eff}}$, $[M/H]_*$, log surface gravity ($\log{g_*}$), and $\log (R_*/D)$, where $D$ is the distance to the system. We calculated the posterior probability distributions of these parameters using the Markov chain Monte Carlo (MCMC) method implemented in the Python package \texttt{emcee} \citep{foreman-mackey2013emcee}. At each MCMC step, a synthetic spectrum was calculated by linearly interpolating the model grid for a given set of parameters. A Gaussian prior from IRD was applied for $[M/H]_*$. A white noise jitter term, $\sigma_{\rm jitter}$, was also fitted for each of the Gaia EDR3, 2MASS, and WISE datasets such that the magnitude uncertainty was given by $\sqrt{\sigma_{\rm cat}^2 + \sigma_{\rm jitter}^2}$, where $\sigma_{\rm cat}$ is the catalogued uncertainty in magnitude. Using the obtained posteriors of $\log (R_*/D)$ and $T_{\rm {*,eff}}$, we also derived the posteriors of $R_*$ and $L_*$ applying the distance from \citet{bailer-jones2021} for $D$, which are based on the Gaia EDR3 parallaxes \citep{lindegren2021}.
Finally, we estimated the stellar mass ($M_*$) and radius from $[Fe/H]_*$ and the absolute $K_s$ magnitude via the empirical relations of \cite{mann2019} and \cite{mann2015}, respectively.

Tables \ref{tab:toi1442_params} and \ref{tab:toi2445_params} report the stellar parameters obtained with the three methods and values adopted from the literature. The parameter values obtained with multiple methods are consistent within 1$\sigma$. The $\log{g_*}$ is weakly constrained by SED fitting. The empirical relations provide the most precise determinations of $M_*$, $\rho_*$, and $\log{g_*}$; therefore, we adopted these values to compute the planetary parameters.

\section{Light curve analysis}
\label{sec:lc_analysis}

\subsection{Contamination transit models}
\label{sec:cont_tran_model}
In order to validate the planetary nature of our TOIs, we fitted simultaneously the TESS and ground-based photometric light curves by modelling planetary transits with  third-light contamination. Our approach is similar to that described by \cite{parviainen2019,parviainen2020,parviainen2021} and \cite{esparza-borges2022}.

The transit models were generated with a customised version of \texttt{PYLIGHTCURVE}\footnote{\url{https://github.com/ucl-exoplanets/pylightcurve}} \citep{tsiaras2016}, which implements the analytic formulae derived by \cite{pal2008}. Conventionally, the model light curves are normalised so that the out-of-transit flux is unity, and the flux drop during transit corresponds to the fraction of stellar flux occulted by the transiting planet. We fitted the following transit parameters: planet-to-star radius ratio ($p=R_{\mathrm{p}}/R_*$), orbital period ($P$), epoch of transit ($T_0$), total transit duration ($T_{14}$), stellar mean density ($\rho_*$), and two limb-darkening coefficients (LDCs, $q_1$ and $q_2$).
We adopted the power-2 law to approximate the stellar limb-darkening profile \citep{hestroffer1997}, as recommended by \cite{morello2017}, especially for M dwarfs. Additionally, we implemented optimal sampling by means of the transformed LDCs, $q_1$ and $q_2$, derived by \cite{short2019} following the procedure of \cite{kipping2013}.

Third light generally means any contribution to the flux from sources outside the star--planet system, such as blended or nearby stars a  part of whose photons hit the selected photometric aperture of the target.
In this work, we define the relative third-light flux as
\begin{equation}
\label{eqn:def_blend}
\beta = \frac{F_{\mathrm{c}}}{F_* + F_{\mathrm{c}}} ,
\end{equation}
where $F_*$ and $F_{\mathrm{c}}$ denote the flux from the target star and contaminating sources, respectively. Thus, the contamination transit model can be expressed as
\begin{equation}
\hat{F}(t) = (1-\beta)(1-\Lambda(t)) + \beta ,
\end{equation}
where $\hat{F}(t)$ is the normalised astrophysical flux, and $1-\Lambda(t)$ is the pure planetary transit model. We further assumed that the contaminating flux comes from only one blended star, except for the TESS observations, due to the significantly larger pixel scale of TESS compared to that of ground-based detectors. The planet self-blend effect is negligible in the analysed datasets \citep{kipping2010,martin-lagarde2020}. We fitted the photospheric parameters of the contaminating star ($T_{\mathrm{c,eff}}$ and $\log{g_\mathrm{c}}$), a flux scaling factor ($f_{\mathrm{c}}$) to account for, for example,  different distances of the target and contaminating stars, and an independent blend TESS factor ($\beta_{TESS}$). The passband-integrated fluxes of the target and hypothetical contaminant stars were computed using \texttt{ExoTETHyS.BOATS}\footnote{\url{https://github.com/ucl-exoplanets/ExoTETHyS}} \citep{morello2021}, based on a precomputed grid of \texttt{PHOENIX} stellar spectra \citep{claret2018}.

\begin{table*}[]
\caption{Prior probability distributions of the fitted parameters.}
\centering
\begin{tabular}{cccc}
\hline
 & Parameter & TOI-1442 & TOI-2445 \\
\hline
\noalign{\smallskip}
Transit & $p$ & $\mathcal{U}(0,1)$ & $\mathcal{U}(0,1)$ \\
 & $P$ ($d$)\tablefootmark{a} & $\mathcal{N}(0.409072,0.000011)$ & $\mathcal{N}(0.371133,0.000047)$ \\
 & $T_0$ ($BTJD$)\tablefootmark{a} & $\mathcal{N}(1683.45058,0.00085)$ & $\mathcal{N}(1411.21732,0.00188)$ \\
 & $\rho_*$ ($\rho_{\odot}$) & $\mathcal{N}(9.50,1.27)$ & $\mathcal{N}(12.45,1.67)$ \\
 & $T_{14}$ ($hr$)\tablefootmark{a} & $\mathcal{U}(0.169,1.036)$ & $\mathcal{U}(0,1.289)$ \\
\hline
\noalign{\smallskip}
Contamination & $T_{\mathrm{*,eff}}$ ($K$)\tablefootmark{b} & 3304 & 3306 \\
 & $\log{g_*}$ (dex)\tablefootmark{c} & 4.91 & 4.96 \\
 & $T_{\mathrm{c,eff}}$ ($K$) & $\mathcal{U}(2300,12000)$ & $\mathcal{U}(2300,12000)$ \\
 & $\log{g_\mathrm{c}}$ (dex) & $\mathcal{U}(2.0,5.5)$ & $\mathcal{U}(2.0,5.5)$ \\
 & $f_{\mathrm{c}}$\tablefootmark{d} & $\mathcal{U}(-1,1000)$ & $\mathcal{U}(-1,1000)$ \\
 & $\beta_{TESS}$\tablefootmark{d} & $\mathcal{U}(-0.2,0.99)$ & $\mathcal{U}(-0.2,0.99)$ \\
\hline
\noalign{\smallskip}
LDCs\tablefootmark{e} & $q_{1,TESS}$ & $\mathcal{N}(0.038667,0.1)\times\mathcal{U}(0,1)$ & $\mathcal{N}(0.038237,0.1)\times\mathcal{U}(0,1)$ \\
 & $q_{2,TESS}$ & $\mathcal{N}(0.005047,0.1)\times\mathcal{U}(0,1)$ & $\mathcal{N}(0.00067,0.1)\times\mathcal{U}(0,1)$ \\
 & $q_{1,g}$ & $\mathcal{N}(0.104994,0.1)\times\mathcal{U}(0,1)$ & $\mathcal{N}(0.104916,0.1)\times\mathcal{U}(0,1)$ \\
 & $q_{2,g}$ & $\mathcal{N}(0.0,0.1)\times\mathcal{U}(0,1)$ & $\mathcal{N}(0.0,0.1)\times\mathcal{U}(0,1)$ \\
 & $q_{1,r}$ & $\mathcal{N}(0.126307,0.1)\times\mathcal{U}(0,1)$ & $\mathcal{N}(0.123358,0.1)\times\mathcal{U}(0,1)$ \\
 & $q_{2,r}$ & $\mathcal{N}(0.102417,0.1)\times\mathcal{U}(0,1)$ & $\mathcal{N}(0.092309,0.1)\times\mathcal{U}(0,1)$ \\
 & $q_{1,i}$ & $\mathcal{N}(0.047087,0.1)\times\mathcal{U}(0,1)$ & $\mathcal{N}(0.047093,0.1)\times\mathcal{U}(0,1)$ \\
 & $q_{2,i}$ & $\mathcal{N}(0.0,0.1)\times\mathcal{U}(0,1)$ & $\mathcal{N}(0.0,0.1)\times\mathcal{U}(0,1)$ \\
 & $q_{1,z_s}$ & $\mathcal{N}(0.031820,0.1)\times\mathcal{U}(0,1)$ & $\mathcal{N}(0.031817,0.1)\times\mathcal{U}(0,1)$ \\
 & $q_{2,z_s}$ & $\mathcal{N}(0.0,0.1)\times\mathcal{U}(0,1)$ & $\mathcal{N}(0.0,0.1)\times\mathcal{U}(0,1)$ \\
 & $q_{1,I}$ & $\mathcal{N}(0.028676,0.1)\times\mathcal{U}(0,1)$ & $\mathcal{N}(0.028667,0.1)\times\mathcal{U}(0,1)$ \\
 & $q_{2,I}$ & $\mathcal{N}(0.0,0.1)\times\mathcal{U}(0,1)$ & $\mathcal{N}(0.0,0.1)\times\mathcal{U}(0,1)$ \\
 & $q_{1,I+z}$ & -- & $\mathcal{N}(0.022097,0.1)\times\mathcal{U}(0,1)$ \\
 & $q_{2,I+z}$ & -- & $\mathcal{N}(0.0,0.1)\times\mathcal{U}(0,1)$ \\
\hline
\noalign{\smallskip}
Baseline & $\log_{10}{\sigma_{\mathrm{GP}}}$ & $\mathcal{U}(-10,6)$ & $\mathcal{U}(-10,6)$ \\
 & $\log_{10}{\rho_{\mathrm{GP}}}$ & $\mathcal{U}(-10,6)$ & $\mathcal{U}(-10,6)$ \\
 & $N_{TESS}$\tablefootmark{f} & $\mathcal{U}(1000,2000)$ & $\mathcal{U}(400,1400)$ \\
 & $N_{ground}$\tablefootmark{f} & $\mathcal{U}(0.5,1.5)$ & $\mathcal{U}(0.5,1.5)$ \\
 & $X_1$, $X_2$\tablefootmark{g} & $\mathcal{U}(-1,1)$ & $\mathcal{U}(-1,1)$ \\
\hline
\end{tabular}
\tablefoot{
$\mathcal{U}(a,b)$ denotes a uniform prior delimited by $a$ and $b$; $\mathcal{N}(\mu,\sigma)$ denotes a normal prior with $\mu$ mean and $\sigma$ width.\\
\tablefoottext{a}{Centred on the ExoFOP values as of 2021 November 30. For the normal distributions, the prior $\sigma$ equals three times the nominal error bars. For the uniform distributions, the whole interval equals six times the nominal error bars.}\\
\tablefoottext{b}{Arithmetic average between spectral synthesis and SED fitting estimates from Tables \ref{tab:toi1442_params} and \ref{tab:toi2445_params}.}\\
\tablefoottext{c}{Obtained with empirical relations from Tables \ref{tab:toi1442_params} and \ref{tab:toi2445_params}.}\\
\tablefoottext{d}{Negative contamination values are allowed to avoid bouncing effect at the physical boundary of zero contaminants; they could also be caused by systematic offsets.}\\
\tablefoottext{e}{Some ground-based detectors share the same LDCs if they have the same passband (see Table \ref{tab:detr_aux_params}).}\\
\tablefoottext{f}{Normalisation factors (i.e. multiplicative factors for the contamination transit models). A factor is fitted for each TESS sector and for each ground-based observation.}\\
\tablefoottext{g}{Coefficients of the linear detrending models for ground-based observation, using up to two auxiliary parameters (see Table \ref{tab:detr_aux_params}).}
}
\label{tab:priors}
\end{table*}

\begin{figure*}
\centering
\includegraphics[width=0.99\hsize]{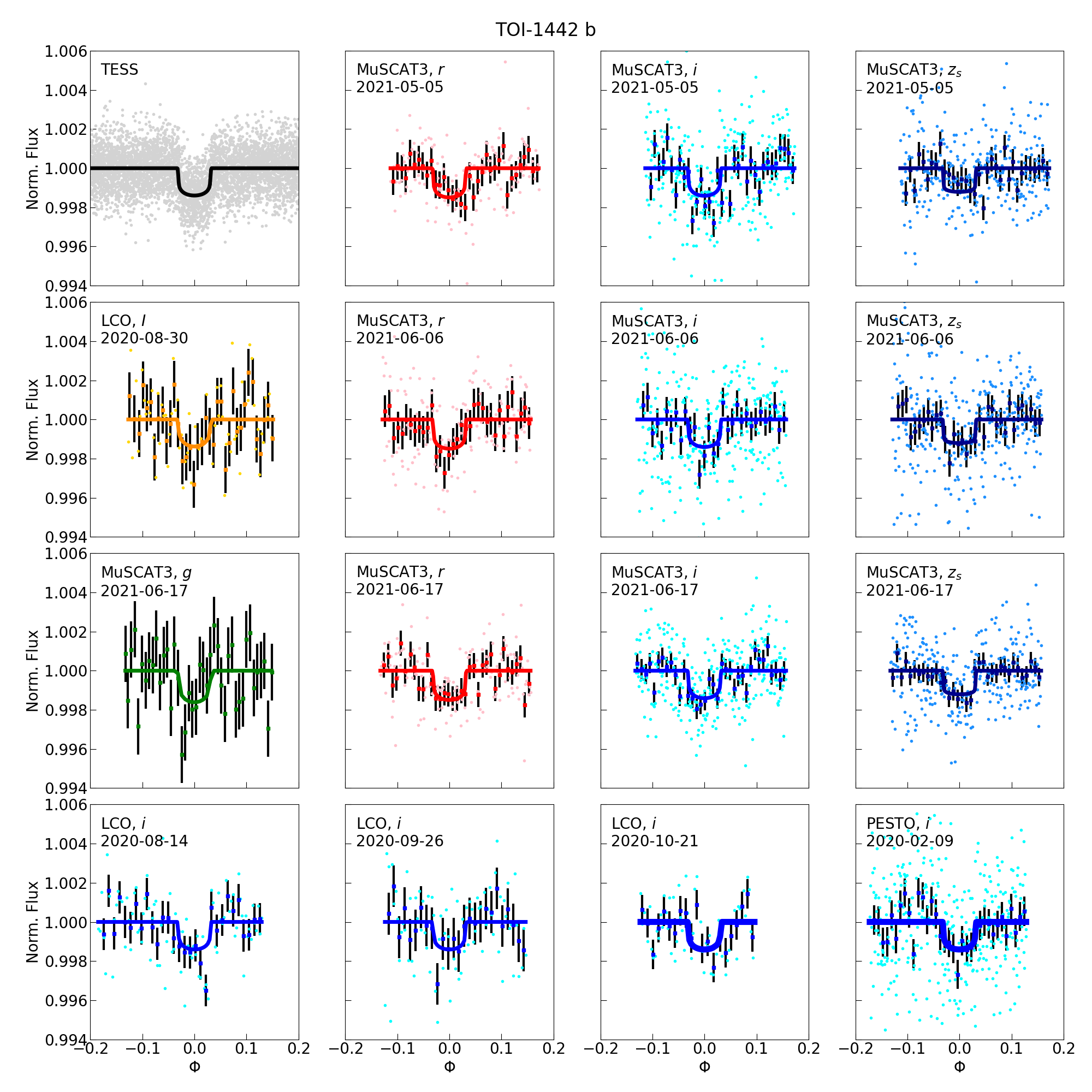}
\caption{Photometric observations of TOI-1442 b. Top-left panel: TESS phase-folded light curve of TOI-1442 b with bin factor of 20 (gray) and best-fitting transit model (black). Other panels: Ground-based light curves after data detrending (lighter dots), with 5-min bins (darker dots) and corresponding error bars, and best-fitting transit models (solid lines).
}
\label{fig:toi1442_detr_lc}
\end{figure*}

\begin{figure*}
\centering
\includegraphics[width=0.99\hsize]{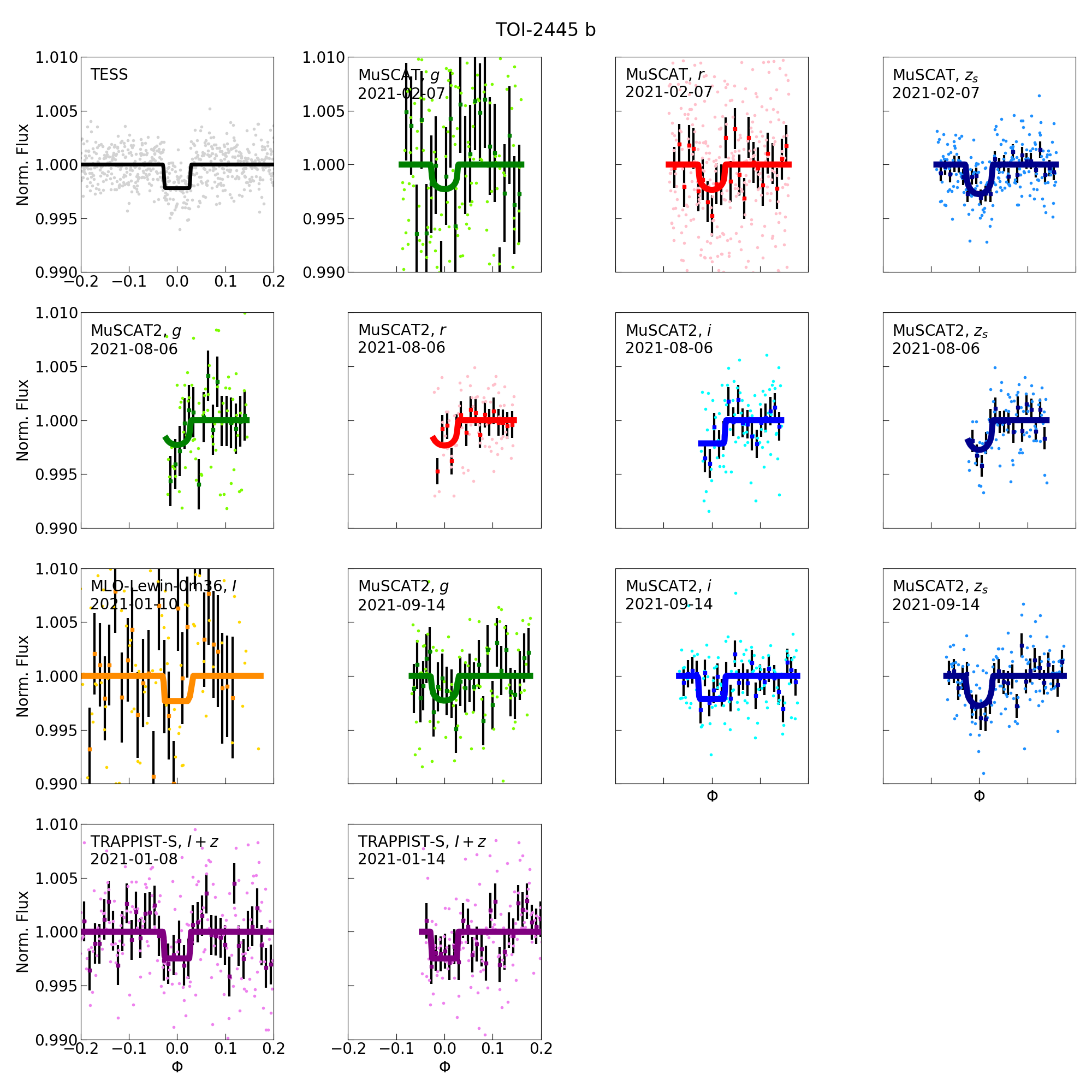}
\caption{TESS phase-folded and ground-based light curves of TOI-2445 b. Analogous to Figure \ref{fig:toi1442_detr_lc}.
}
\label{fig:toi2445_detr_lc}
\end{figure*}

\subsection{Baseline models}
\label{sec:baselines}
In addition to planetary transit and possible third-light contamination, other signals are present in the observed light curves,  of astrophysical and instrumental origin. We modelled the modulations present in the TESS time series by using Gaussian processes (GPs), which provide a flexible non-parametric method to approximate stochastic trends in various types of datasets \citep{rasmussen2006gp,roberts2012}. In the field of astrophysics, GPs are often used to filter out stellar variability and instrumental systematic effects in photometric time series (e.g. \citealp{gibson2012,evans2015,barros2020}). In this paper we computed the TESS GPs using \texttt{celerite}\footnote{\url{https://github.com/dfm/celerite}} \citep{foreman-mackey2017celerite} with the Matern-3/2 kernel:
\begin{equation}
k(\tau) = \sigma_{\mathrm{GP}}^2 \left [ \left (1 + \frac{1}{\epsilon} \right ) e^{-(1-\epsilon) \sqrt{3} \tau/ \rho_{\mathrm{GP}}} + \left (1 - \frac{1}{\epsilon} \right ) e^{-(1+\epsilon) \sqrt{3} \tau/ \rho_{\mathrm{GP}}} \right ] .
\end{equation}
Here $\tau=|t_i - t_j|$ is the time interval between two points, $\epsilon = 0.01$, and  $\sigma_{\mathrm{GP}}$ and $\rho_{\mathrm{GP}}$ are the characteristic amplitude and timescale of the modulations. The choice of GP Matern-3/2 kernel has proven effective to detrend TESS photometry (e.g. \citealp{gonzalez-alvarez2021,kossakowski2021,murgas2021}).

We detrended the ground-based light curves by fitting linear models with maximum two auxiliary parameters, as recommended in the data reports uploaded by the relevant observing teams on the ExoFOP website. The last column of Table \ref{tab:detr_aux_params} lists the names of the auxiliary parameters used for detrending, as they are given in the original reports.

\subsection{Bayesian priors}

We made use of the \texttt{emcee} package to sample the posterior probability distributions of the astrophysical parameters associated with the contamination transit models (Section \ref{sec:cont_tran_model}) and with the baseline model parameters (Section \ref{sec:baselines}), simultaneously. The adopted prior distributions were generally broader than the potential constraints available from ancillary observations (e.g. the stellar spectra). They were chosen to discard unphysical solutions without biasing or boosting the inferences from the colour contamination analysis. Table \ref{tab:priors} lists the Bayesian priors for all parameters.

We selected uniform priors on the radius ratio and total transit duration to let them be fully constrained by the transit light curves themselves. These parameters were shared by all observations and passbands; second-order effects such as the wavelength-dependent absorption of the planetary atmospheres and possible changes in orbital inclinations were neglected. We also assumed linear ephemerides with Gaussian priors on the orbital period and epoch of transits, centred on the ExoFOP values, but their $\sigma$ widths were conservatively enhanced by a factor of 3. For the stellar mean density, we used our estimates from empirical relations (Tables \ref{tab:toi1442_params} and \ref{tab:toi2445_params}) with a factor of 1.5 on the error bars. The orbital eccentricity was fixed to zero, as  expected for USP planets.

The stellar LDCs were computed with \texttt{ExoTETHyS.SAIL} \citep{morello2020,morello2020joss}, based on the photospheric parameters  in Tables \ref{tab:toi1442_params} and \ref{tab:toi2445_params}, then transformed into $q_1$ and $q_2$ (see Section \ref{sec:cont_tran_model}). We adopted broad Gaussian priors centred on the theoretical values of $q_1$ and $q_2$ with widths $\sigma=0.1$, that largely encompass all plausible values derived from stellar models with consistent photospheric parameters. 

To compute the physical contamination model, we fixed the effective temperature and surface gravity of the target star, and assumed uniform priors for all the contaminant parameters, including the blend TESS factor. The choice of uniform priors is a conservative one as it constrains the possible flux contamination based only on the photometric time series. The previous knowledge of the field around the target stars (see Section \ref{sec:obs_tess} and Figure \ref{fig:tpf_plots}), as well as the high resolution images analysed by \cite{giacalone2022}, point towards a low probability and/or amount of blending between sources.

Finally, we selected uninformative prior distributions for all data detrending parameters and normalisation factors. In particular, we adopted log-uniform priors for the TESS GP parameters, that were shared over all sectors for the same target. We adopted uniform priors for the coefficients of the linear detrending models of the ground-based observations and for the normalisation factors. All data detrending and astrophysical parameters were fitted simultaneously.
The final \texttt{emcee} fit for TOI-1442 had 68 parameters, 300 walkers, and 300\,000 iterations. 
The final \texttt{emcee} fit for TOI-2445 had 48 parameters, 200 walkers, and 300\,000 iterations.
We applied a conservative burn-in of 100\,000 iterations, which is much longer than the autocorrelation lengths of each parameter chain.
Figures \ref{fig:toi1442_detr_lc} and \ref{fig:toi2445_detr_lc} show the TESS phase-folded and ground-based light curves, after data detrending, and maximum likelihood contamination transit models for TOI-1442 b and TOI-2445 b, respectively.

\begin{table*}[]
\caption{Final system parameters}
\centering
\begin{tabular}{cccc}
\hline
 & Parameter & TOI-1442 & TOI-2445 \\
\hline
\noalign{\smallskip}
Stellar & $T_{\mathrm{*,eff}}$ ($K$) & $3304 \pm 100$ & $3306 \pm 100$ \\
 & $\log{g_*}$ (dex) & $4.91 \pm 0.03$ & $4.96 \pm 0.03$ \\
 & $[M/H]_*$ (dex) & $0.11 \pm 0.07$ & $-0.33\pm0.07$ \\
 & $[Fe/H]_*$ (dex) & $0.03\pm0.15$ & $-0.39\pm0.16$ \\
 & $M_*$ ($M_{\odot}$) & $0.2843\pm0.0065$ & $0.2448\pm0.0056$ \\
 & $R_*$ ($R_{\odot}$) & $0.3105\pm0.0089$ & $0.2699\pm0.0078$ \\
 & $L_*$ ($L_{\odot}$) & $0.01020_{-0.00012}^{+0.00014}$ & $0.00760_{-0.00017}^{+0.00019}$ \\
\noalign{\smallskip}
\hline
Transit fit & $p$ & $0.03405 \pm 0.00077$ & $0.0453 \pm 0.0018$ \\
\noalign{\smallskip}
 & $P$ ($d$) & $0.4090682 \pm 0.0000004$ & $0.3711286 \pm 0.0000004$ \\
\noalign{\smallskip}
 & $T_0$ ($BTJD$) & $1683.45193_{-0.00034}^{+0.00033}$ & $1411.21990_{-0.00083}^{+0.00071}$ \\
\noalign{\smallskip}
 & $\rho_*$ ($\rho_{\odot}$) & $9.09_{-1.06}^{+0.91}$ & $12.5 \pm 1.6$ \\
\noalign{\smallskip}
 & $T_{14}$ ($hr$) & $0.644_{-0.013}^{+0.016}$ & $0.537_{-0.019}^{+0.022}$ \\
\noalign{\smallskip}
(derived) & $a/R_*$\tablefootmark{a} & $4.84_{-0.20}^{+0.16}$ & $5.04_{-0.23}^{+0.20}$ \\
\noalign{\smallskip}
 & $b$\tablefootmark{b} & $0.29 \pm 0.13$ & $0.44_{-0.13}^{+0.09}$ \\
\noalign{\smallskip}
 & $i$ ($deg$)\tablefootmark{c} & $86.57_{-1.62}^{+1.54}$ & $84.95_{-1.30}^{+1.53}$ \\
\noalign{\smallskip}
\hline
\noalign{\smallskip}
RVs & $K_{\mathrm{p}}$ ($\mathrm{m \, s}^{-1}$) & $<34$ & $<82$ \\
\noalign{\smallskip}
\hline
\noalign{\smallskip}
Planetary and orbital & $R_{\mathrm{p}}$ ($R_{\oplus}$) & $1.15 \pm 0.06$ & $1.33 \pm 0.09$ \\
\noalign{\smallskip}
(derived) & $M_{\mathrm{p}}$ ($M_{\oplus}$) & $<8$ & $<18$ \\
\noalign{\smallskip}
 & $a$ ($au$)\tablefootmark{d} & $0.00699_{-0.00049}^{+0.00043}$ & $0.00634_{-0.00047}^{+0.00043}$ \\
\noalign{\smallskip}
 & $T_{\mathrm{p,eq}}$ ($K$)\tablefootmark{e} & $1357_{-42}^{+49}$ & $1330_{-56}^{+61}$ \\
\noalign{\smallskip}
\hline
\end{tabular}
\tablefoot{Values preceded by $<$ report 3$\sigma$ upper limits.\\
\tablefoottext{a}{Orbital semi-major axis relative in units of the stellar radius.}\\
\tablefoottext{b}{Impact parameter.}\\
\tablefoottext{c}{Orbital inclination.}\\
\tablefoottext{d}{Orbital semi-major axis.}\\
\tablefoottext{e}{Dayside equilibrium temperature, assuming zero albedo and no heat redistribution.}
}
\label{tab:final_params}
\end{table*}

\begin{figure*}
\centering
\includegraphics[width=0.99\hsize]{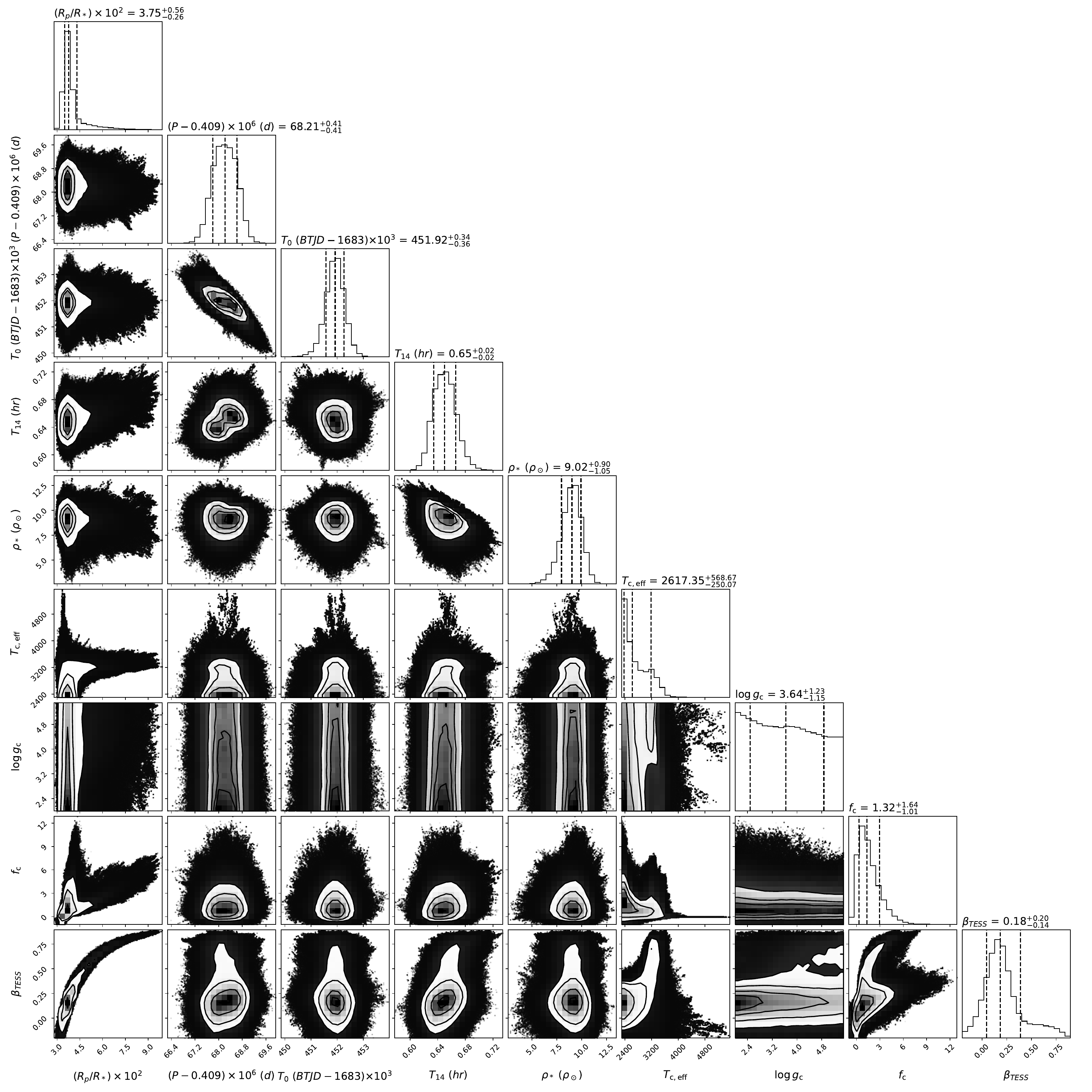}
\caption{Cornerplot showing the posterior distributions and mutual correlations of the transit model parameters for TOI-1442. The histograms along the diagonal give the median value 1$\sigma$ error bars (absolute differences between the medians and the 16th and 84th quantiles) of the distributions. This plot was generated by using the \texttt{corner} Python package \citep{foreman-mackey2016corner}.
}
\label{fig:toi1442_cornerplot}
\end{figure*}

\begin{figure*}
\centering
\includegraphics[width=0.99\hsize]{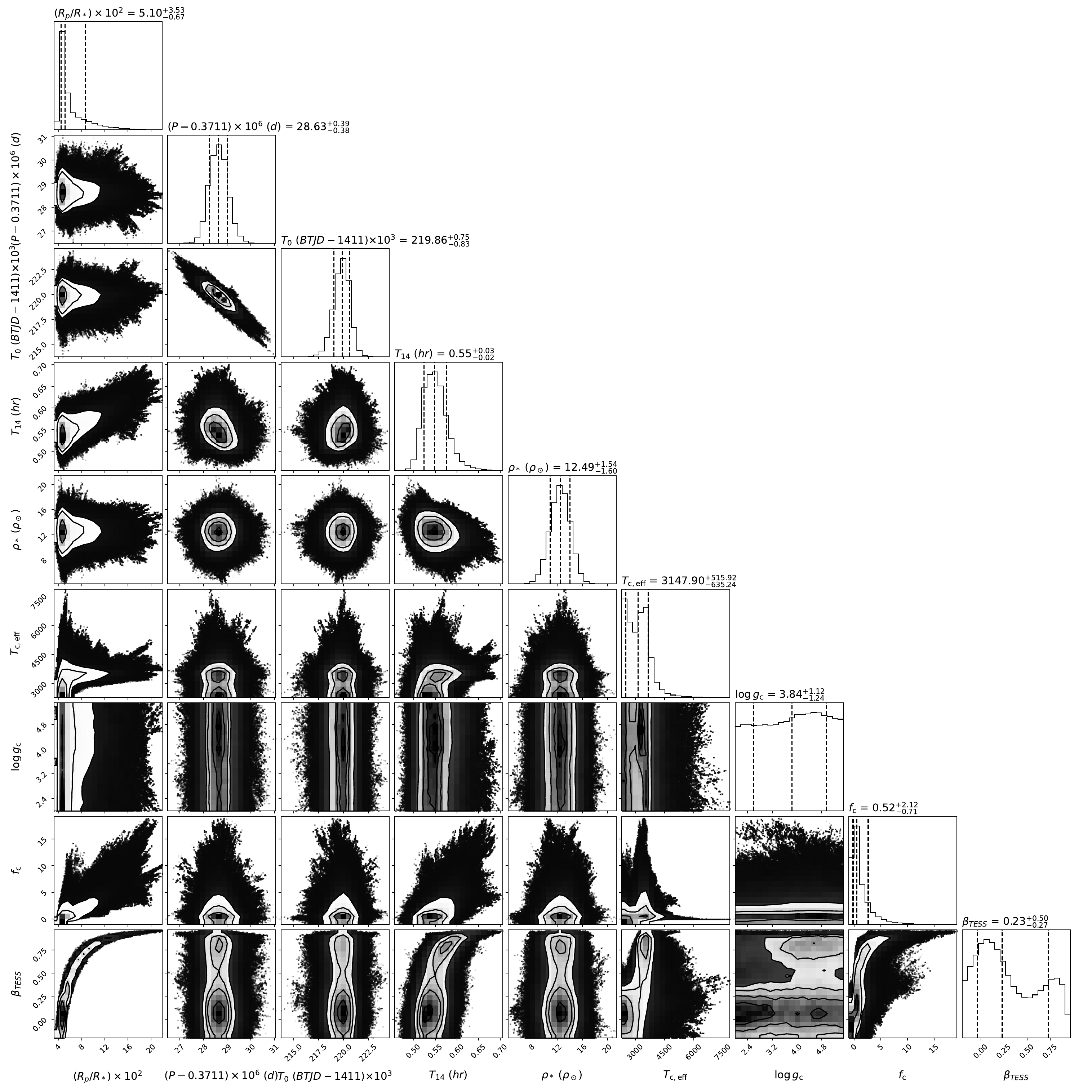}
\caption{Cornerplot showing the posterior distributions and mutual correlations of the transit model parameters for TOI-2445 (analogous to Figure \ref{fig:toi1442_cornerplot}).
}
\label{fig:toi2445_cornerplot}
\end{figure*}

\begin{table*}[]
\caption{Best-fit transit and contamination parameters.}
\centering
\begin{tabular}{cccc}
\hline
 & Parameter & TOI-1442 & TOI-2445 \\
\hline
\noalign{\smallskip}
Transit (fitted) & $p$ & $0.0375_{-0.0026}^{+0.0056}$ & $0.051_{-0.007}^{+0.035}$ \\
\noalign{\smallskip}
 & $P$ ($d$) & $0.4090682 \pm 0.0000004$ & $0.3711286 \pm 0.0000004$ \\
\noalign{\smallskip}
 & $T_0$ ($BTJD$) & $1683.45192_{-0.00036}^{+0.00034}$ & $1411.21986_{-0.00083}^{+0.00075}$ \\
\noalign{\smallskip}
 & $\rho_*$ ($\rho_{\odot}$) & $9.02_{-1.05}^{+0.90}$ & $12.5 \pm 1.6$ \\
\noalign{\smallskip}
 & $T_{14}$ ($hr$) & $0.650 \pm 0.016$ & $0.548_{-0.024}^{+0.028}$ \\
\noalign{\smallskip}
Contaminant (fitted) & $T_{\mathrm{c,eff}}$ & $<4100$ & $<5900$ \\
\noalign{\smallskip}
 & $\log{g_\mathrm{c}}$ & unconstrained & unconstrained \\
\noalign{\smallskip}
 & $f_{\mathrm{c}}$ & $1.32_{-1.01}^{+1.64}$ & $0.52_{-0.71}^{+2.12}$ \\
\noalign{\smallskip}
 & $\beta_{TESS}$ & $0.18_{-0.14}^{+0.20}$ & $0.23_{-0.27}^{+0.50}$ \\
\noalign{\smallskip}
Transit (derived) & $a/R_*$ & $4.83_{-0.19}^{+0.16}$ & $5.04_{-0.23}^{+0.20}$ \\
\noalign{\smallskip}
 & $b$ & $0.29_{-0.13}^{+0.12}$ & $0.44_{-0.13}^{+0.10}$ \\
\noalign{\smallskip}
 & $i$ ($deg$) & $86.56_{-1.67}^{+1.57}$ & $85.00_{-1.33}^{+1.56}$ \\
\noalign{\smallskip}
Contaminant (derived) & $\beta_g-\beta_{z_s}$ & $-0.12_{-0.10}^{+0.11}$ & $0.01_{-0.11}^{+0.10}$ \\
\noalign{\smallskip}
 & $\beta_r-\beta_{z_s}$ & $-0.12_{-0.09}^{+0.10}$ & $0.01_{-0.11}^{+0.10}$ \\
\noalign{\smallskip}
 & $\beta_i-\beta_{z_s}$ & $-0.04 \pm 0.04$ & $0.00 \pm 0.03$ \\
\noalign{\smallskip}
 & $\beta_I-\beta_{z_s}$ & $0.019_{-0.017}^{+0.026}$ & $0.001_{-0.007}^{+0.013}$ \\
\noalign{\smallskip}
 & $\beta_{TRAPPIST}-\beta_{z_s}$ & -- & $0.00_{-0.03}^{+0.04}$ \\
 \noalign{\smallskip}
 & $\beta_{TESS}-\beta_{z_s}$ & $-0.04_{-0.07}^{+0.06}$ & $0.02_{-0.09}^{+0.11}$ \\
\noalign{\smallskip}
\hline
\end{tabular}
\tablefoot{Values preceded by $<$ report 3$\sigma$ upper limits.}
\label{tab:ctran_params}
\end{table*}

\section{Spectral data analysis}
\label{sec:spectral}

\subsection{Cross-correlation functions}
\label{sec:CCF_method}

\begin{figure*}
\centering
\includegraphics[width=0.49\hsize]{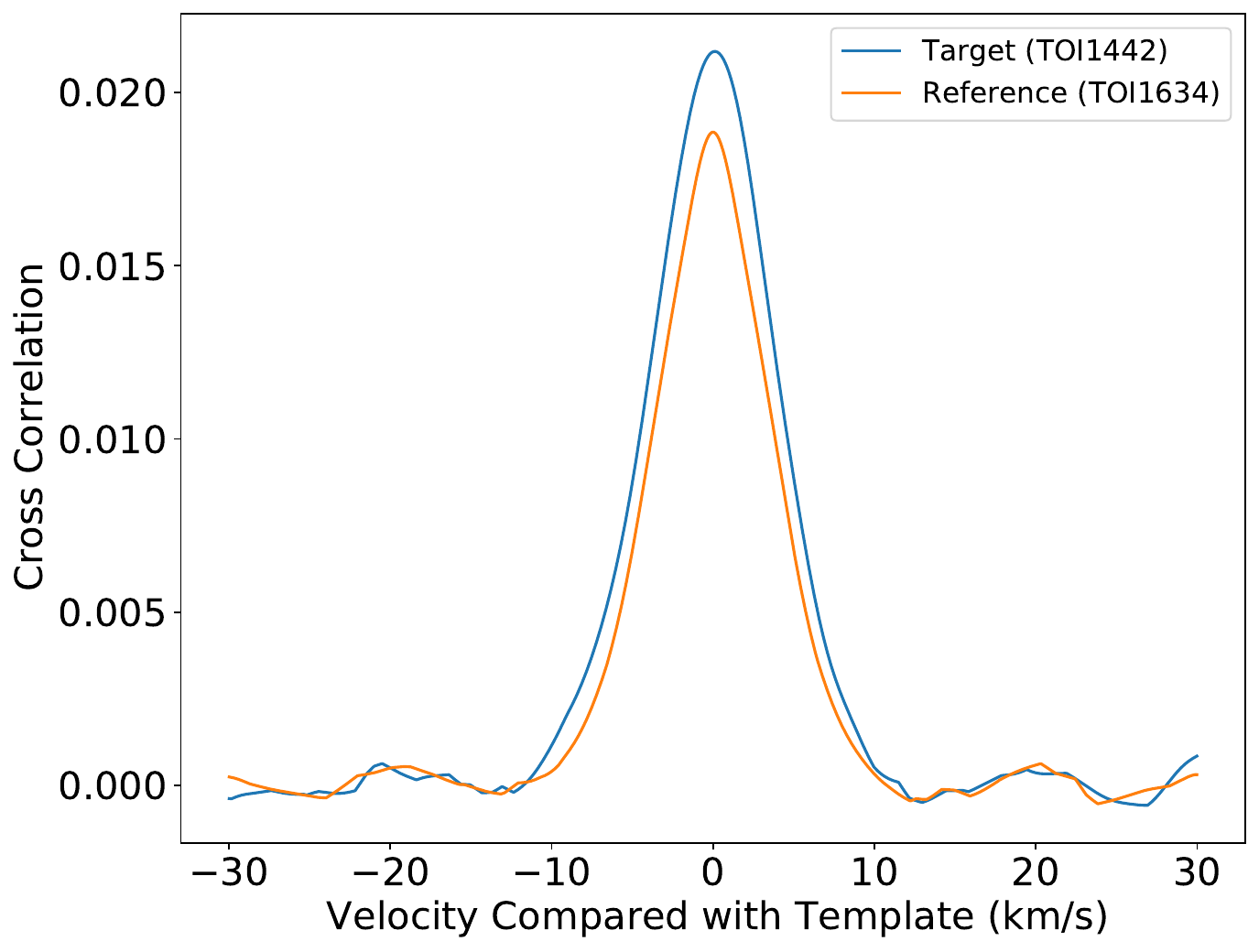}
\includegraphics[width=0.49\hsize]{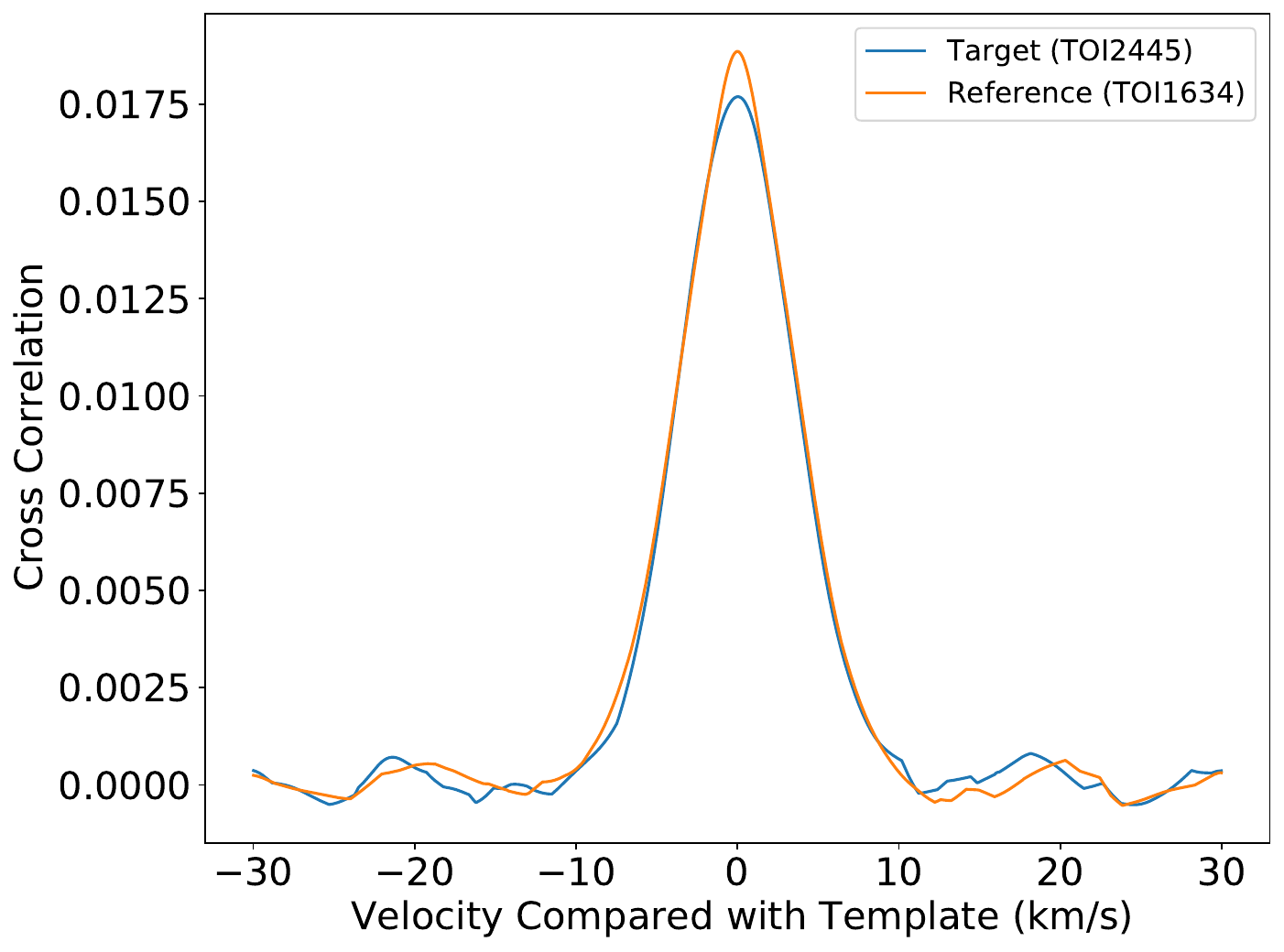}
\caption{Examples of CCFs. Left panel: Calculated CCF of the IRD spectrum of TOI-1442 (blue) taken on UT 2021 June 25 at the orbital phase 0.33, to the template spectrum of TOI-1634, exhibiting a single peak with width of $\sim 10 \, \mathrm{km \, s}^{-1}$. The autocorrelation function of the TOI-1634 spectrum is overplotted as a reference. Right panel: Analogous plot with the CCF of the IRD spectrum of TOI-2445 (blue) taken on UT 2021 October 27 at the orbital phase 0.75.
}
\label{fig:CCFs}
\end{figure*}

We computed the CCFs of IRD spectra of TOI-1442 and TOI-2445 with that of the star TOI-1634, to check whether our targets are double-lined spectroscopic binaries. We selected TOI-1634 as a spectral template as it is a well-studied star \citep{hirano2021}  that shows single-peaked spectra, and it is of similar spectral type ($\sim$ M3 dwarf).
We followed the same method to compute the CCFs, as described by \cite{mori2022} in their Section 3.2. In short, we first computed the CCFs for five spectral segments that are less affected by telluric absorption (i.e. 988-993, 995-1000, 1009-1014, 1016-1021, and 1023-1028 nm), and  then  took their median.

If there are any nearby sources, they could cause two peaks or broadening in the CCFs, if the flux of both stars are detectable. Another possible scenario is a bright source A with a fainter physically bound eclipsing binary BC; the CCFs from both A and BC would show up as a single peak. In this case the fainter CCF of BC would wobble by several km/s, but will not be noticeable in the composed CCF, which would only wobble by a   few m/s. However, this scenario could be unraveled by the correlation between the FWHM of the CCF and the RV.

\subsection{Radial velocities}
\label{sec:RV_method}

\begin{figure*}
\centering
\includegraphics[width=0.49\hsize]{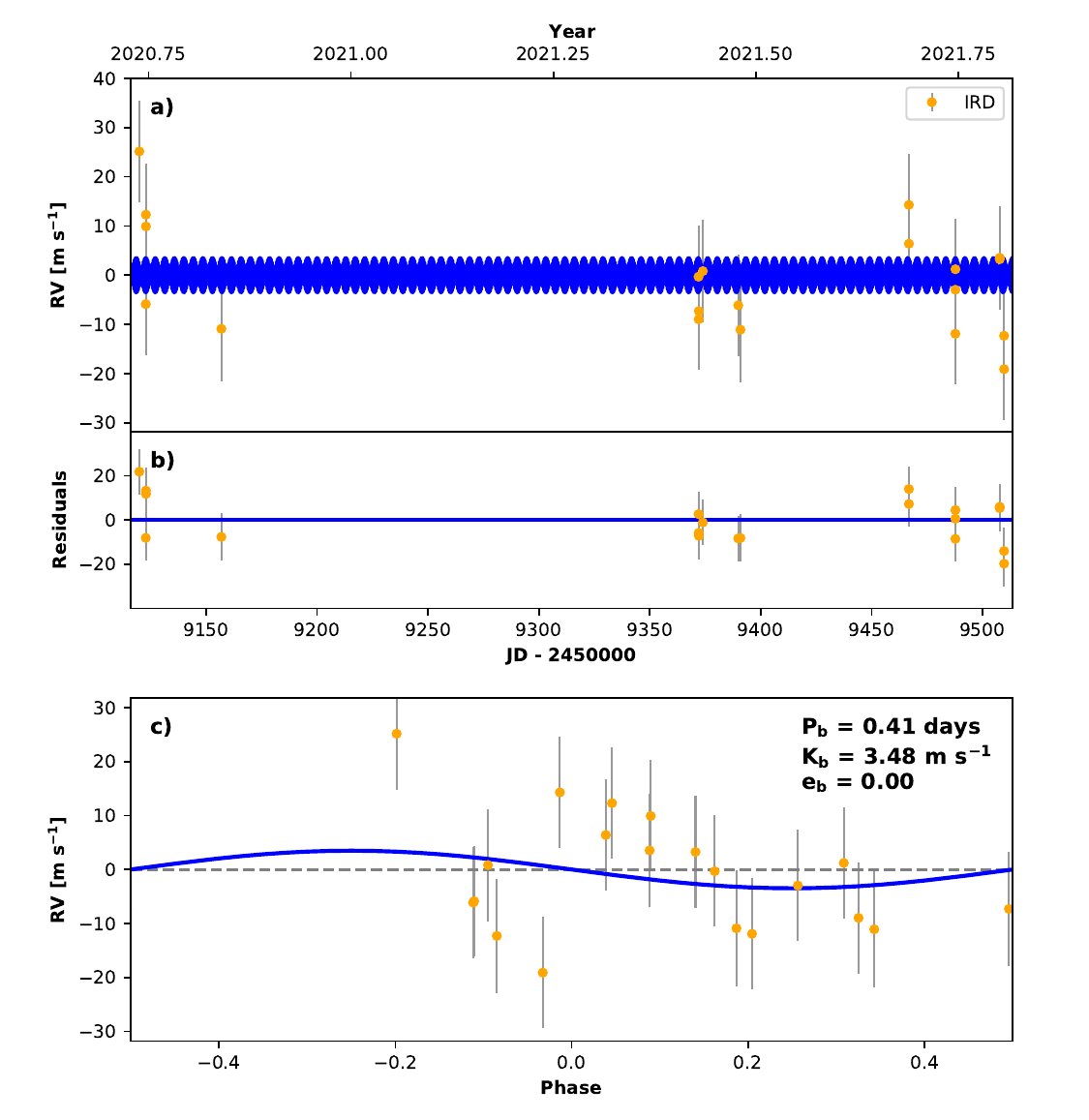}
\includegraphics[width=0.49\hsize]{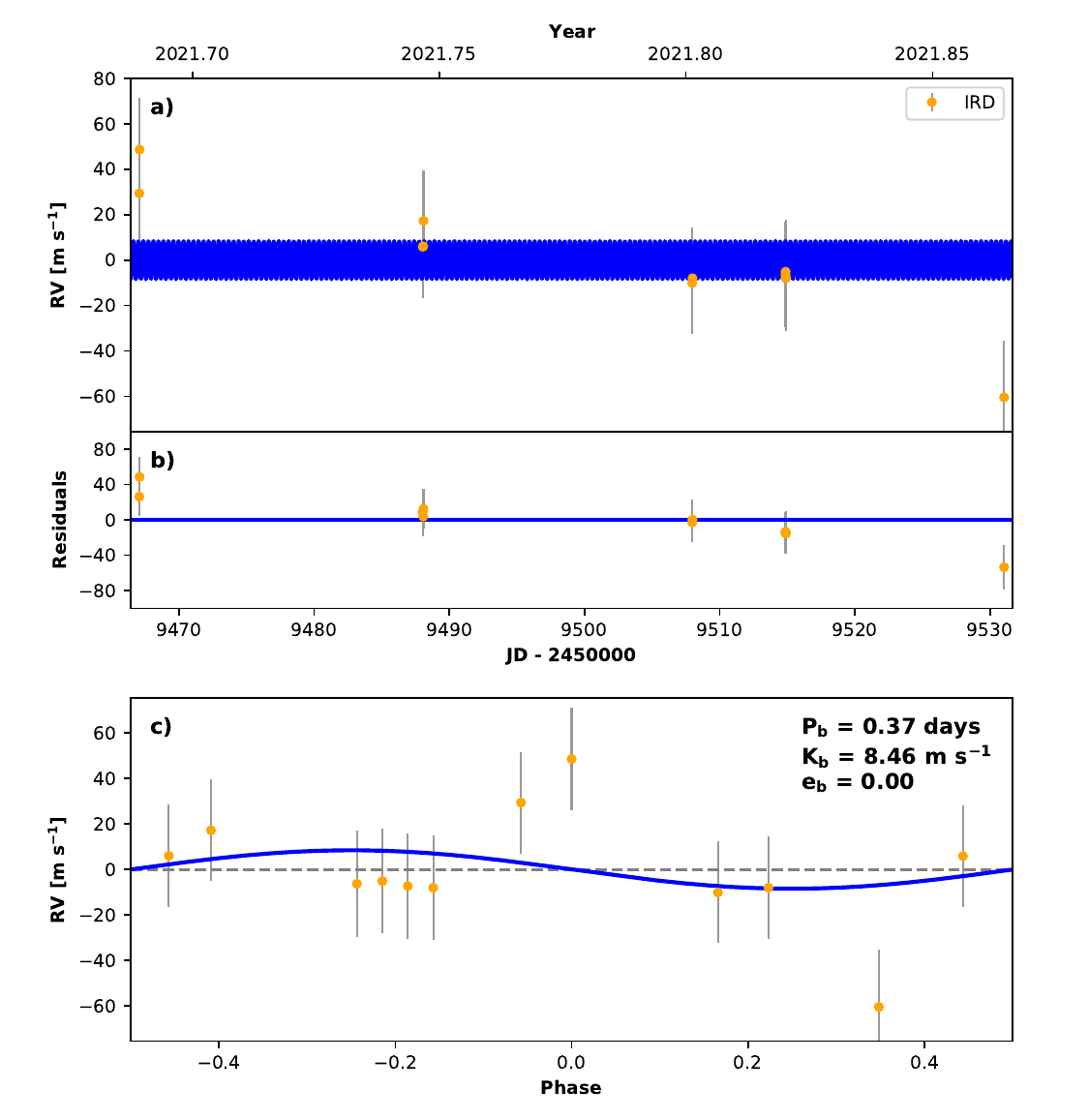}
\caption{Radial velocity data and maximum likelihood \texttt{radvel} models for TOI-1442 (left) and TOI-2445 (right). The top panels display the full timeline data, while the bottom panels show the phase-folded data using the best-fit orbital parameters. The error bars do not account for systematic offsets.
}
\label{fig:radvel_panels}
\end{figure*}

We fitted the IRD RV data using \texttt{radvel}\footnote{\url{https://github.com/California-Planet-Search/radvel}} \citep{fulton2018radvel}. We adopted Gaussian priors on the orbital period and epoch of transits, based on the results of the photometric analysis reported in Table \ref{tab:final_params}, and uniform prior with upper bound of $300 \, \mathrm{m \, s}^{-1}$ on the RV semi-amplitude ($K_{\mathrm{p}}$). We also included the jitter term with uniform prior between 0 and 100.

\section{Results}
\label{sec:results}

\subsection{From the light curve analysis}
Figures \ref{fig:toi1442_cornerplot} and \ref{fig:toi2445_cornerplot} show the posterior distributions and mutual correlations of the contamination transit model parameters for TOI-1442 b and TOI-2445 b, respectively. Table \ref{tab:ctran_params} lists the corresponding median and 1$\sigma$ intervals of the posteriors and those of the derived parameters. We cannot completely rule out some degree of contamination, but we do limit contamination from sources with a colour difference. These partial inferences are expected due to the low S/N of the observed light curves \citep{parviainen2019}. In particular, we note that the differential third-light fraction
\begin{equation}
\Delta \beta_{pass} = \beta_{pass} - \beta_{z_s} ,
\end{equation}
with $\beta$ defined as in Equation \ref{eqn:def_blend}, is consistent with zero within 1.3$\sigma$ for all passbands. Equivalently, the $T_{\mathrm{c,eff}}$ of a hypothetical blended source must have similar or cooler values than $T_{\mathrm{*,eff}}$. The posterior distributions of $T_{\mathrm{c,eff}}$ are bimodal with a peak towards the lower temperature limit of the \texttt{PHOENIX} stellar models grid ($2300 \, K$), and a second peak at $T_{\mathrm{c,eff}} \sim 3000 \, K$. Overall, we pose 3$\sigma$ upper limits at $T_{\mathrm{c,eff}} < 4100 \, K$ and $5900 \, K$ to TOI-1442 b and TOI-2445 b, respectively.

The radius ratios turned out to be $p=0.0375_{-0.0026}^{+0.0056}$ for TOI-1442 b and $p=0.0510_{-0.0067}^{+0.0353}$ for TOI-2445 b. Multiplying these values by $R_*$ from Tables \ref{tab:toi1442_params} and \ref{tab:toi2445_params}, the radii of the planetary candidates are $R_{\mathrm{p}} =1.27_{-0.12}^{+0.23} \, R_{\oplus}$ and $R_{\mathrm{p}} =1.50_{-0.24}^{+1.08} \, R_{\oplus}$, respectively. We also report 3$\sigma$ upper limits of $R_{\mathrm{p}}<3.22 R_{\oplus}$ and $R_{\mathrm{p}}<5.97 R_{\oplus}$.
The 3$\sigma$ upper limits on the radii are smaller than the minimum radius of a brown dwarf \citep{burrows2001}, even in the unlikely cases of strong third-light contamination. The corresponding false alarm probabilities are $<10^{-6}$ (TOI-1442 b) and $1.6 \times 10^{-4}$ (TOI-2445 b), thus validating the planetary nature of both transiting objects.

The radius ratios are largely degenerate with the third-light contamination parameters, leading to heavy tails on the right side of the posterior distributions. We repeated the light curve fits assuming that the third-light contamination is negligible to measure the planetary radii with greater precision. This assumption is supported by various pieces of evidence coming from the above colour contamination analysis, the spectral CCFs analysis (see Sections \ref{sec:CCF_method} and \ref{sec:results_CCFs}), the Gaia DR2 (see Figure \ref{fig:tpf_plots}) and high-resolution imaging \citep{giacalone2022}. We obtained $p=0.03405 \pm 0.00077$ and $R_{\mathrm{p}} =1.15 \pm 0.06 \, R_{\oplus}$ for TOI-1442 b, and $p=0.0453 \pm 0.0018$ and $R_{\mathrm{p}} =1.33 \pm 0.09 \, R_{\oplus}$ for TOI-2445 b. According to the classification of small planets proposed by \cite{luque2022}, TOI-1442 b and TOI-2445 b belong to the rocky planet population.

We used \texttt{Forecaster}\footnote{\url{https://github.com/chenjj2/forecaster}} to predict the planetary masses, obtaining $M_{\mathrm{p}} =1.56_{-0.52}^{+1.07} \, M_{\oplus}$ (for TOI-1442 b) and $M_{\mathrm{p}} =2.33_{-0.80}^{+1.76} \, M_{\oplus}$ (for TOI-2445 b), based on the probabilistic mass-radius relations from \cite{chen2017}. The corresponding RV amplitudes are $K_{\mathrm{p}}=3.1_{-1.0}^{+2.1} \, \mathrm{m \, s}^{-1}$ and $K_{\mathrm{p}}=5.3_{-1.8}^{+4.0} \, \mathrm{m \, s}^{-1}$. We also determined physical 3$\sigma$ upper limits of $M_{\mathrm{p}} < 8 \, M_{\oplus}$ and $18 \, M_{\oplus}$ respectively, based on the pure iron composition model of \cite{zeng2013}.

We estimated the equilibrium temperatures to be $T_{\mathrm{p,eq}} = 1357_{-42}^{+49} \, K$ (TOI-1442 b) and $1330_{-56}^{+61} \, K$ (TOI-2445 b), assuming zero albedo and no heat redistribution. The equilibrium temperatures are above the $880 \, K$ limit to melt rocks and metals on the surface of the dayside hemisphere \citep{mcarthur2004}.

\subsection{From the spectral CCFs}
\label{sec:results_CCFs}

We did not find any suspicious secondary peaks in any of the spectral CCFs analysed in Section \ref{sec:CCF_method} (see also Figure \ref{fig:CCFs}). Thus, we conclude that there is a small possibility that the targets have   blended sources, or that they must be significantly fainter than the target stars.
We also did not find statistically significant correlations between the FWHM of the CCFs and the RVs, although the sample sizes may be too small for this test. With the available data there is no evidence for a bright source A with a fainter physically bound eclipsing binary BC.

\subsection{From the RVs}

The \texttt{radvel} fits failed to provide a good match to the observed RV data, as  can be seen in Figure \ref{fig:radvel_panels}. The maximum likelihood Keplerian models are not statistically favoured over flat lines, but the standard deviation of the residuals are $\sim$2.6 times (TOI-1442 b) and 3.8 times (TOI-2445 b) larger than the nominal error bars. We also checked that alternative configurations with a linear trend do not provide significant improvements.

There are at least three possible explanations for such a poor agreement between the RV data and our simple RV models. Firstly,  the RV measurements can be dominated by random noise and systematic effects as their overall ranges of variations are comparable with the potential instrumental offsets of $\sim 10\,\mathrm{m \, s}^{-1}$ \citep{delrez2022,mori2022}. Secondly, strong stellar activity could also cause similar offsets. Another possibility is that TOI-1442 and/or TOI-2445 host additional planets with detectable RV signals, but the current data are insufficient to support this scenario.

Nonetheless, the current RV measurements have a good phase coverage, so that we can constrain $K_{\mathrm{p}}$. From the \texttt{radvel} MCMC fits, we derive 3$\sigma$ upper limits of $K_{\mathrm{p}} < 15 \, \mathrm{m \, s}^{-1}$ for TOI-1442 b and $K_{\mathrm{p}} < 43 \, \mathrm{m \, s}^{-1}$ for TOI-2445 b. Using $M_*$ from Tables \ref{tab:toi1442_params} and \ref{tab:toi2445_params}, we infer 3$\sigma$ upper limits on the (projected) planet masses of $M_{\mathrm{p}} \sin{i} \approx M_{\mathrm{p}} < 8 M_{\oplus}$ and $20 M_{\oplus}$, respectively. These mass upper limits are more than $ 100$ times smaller than the accepted minimum mass of a brown dwarf \citep{burrows1997}, thus suggesting the planetary nature of the transiting companions.

\section{Discussion}
\label{sec:discussion}

\begin{figure*}[!h]
\centering
\includegraphics[width=0.9\hsize]{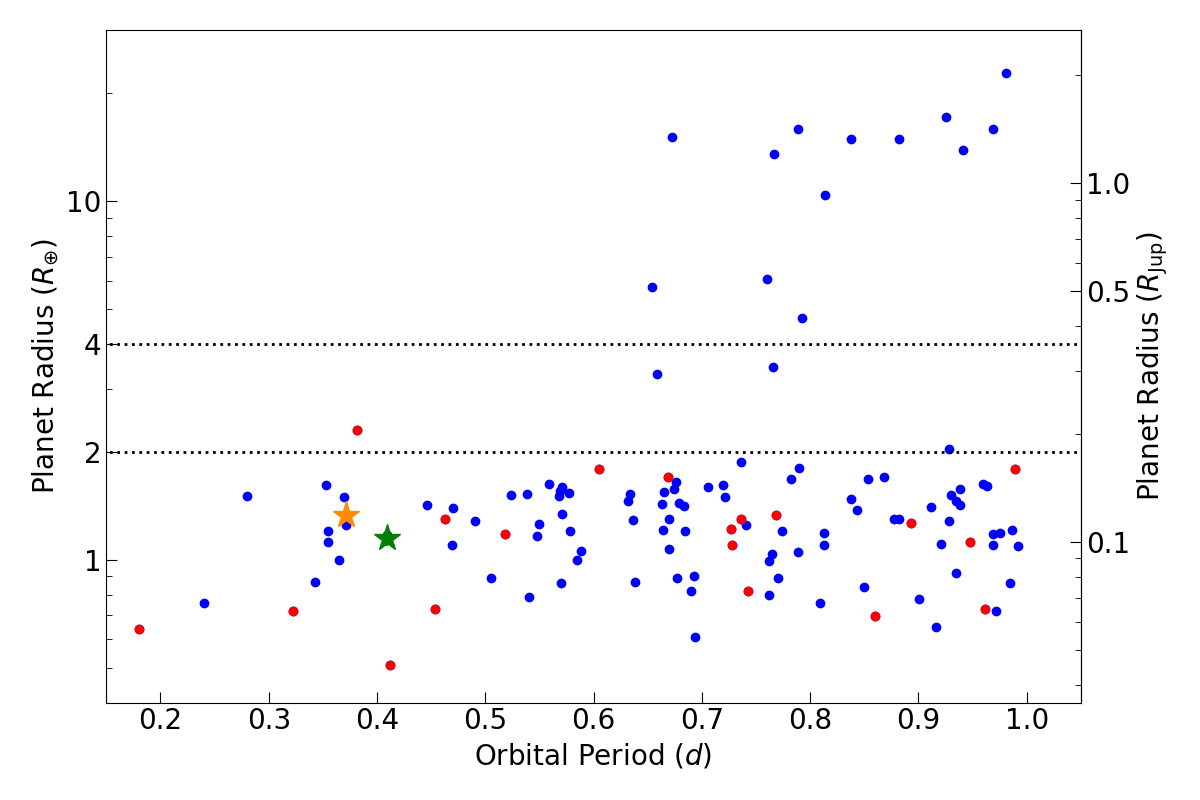}
\caption{Planetary radius vs orbital period for the known USP planets, based on NASA Exoplanet Archive data as of 2022 November 17. Planets around M dwarfs are colored in red. The green and orange stars correspond to TOI-1442 b and TOI-2445 b, based on the final results of our analysis. The horizontal dashed lines delimit the regions $R_{\mathrm{p}} < 2 R_{\oplus}$, encompassing 80$\%$ of the USP population, and $R_{\mathrm{p}} < 4 R_{\oplus}$, which approximates the sub-Neptunes.  
}
\label{fig:P_vs_Rp}
\end{figure*}

\subsection{Comparison with previously published parameters}
Table \ref{tab:final_params} reports our final system parameters. Our stellar, orbital, and planetary parameters are consistent within 1$\sigma$ with those published by \cite{giacalone2022}. The only apparent discrepancy is an offset of nearly +300\,$K$ on the planetary equilibrium temperatures from Table \ref{tab:final_params} compared to those reported by \cite{giacalone2022}. However, this is actually a matter of different definitions adopted. The previous study assumed full atmospheric circulation efficiency. Instead, we calculated the dayside temperature assuming no heat redistribution. This second scenario is closer to what is inferred from the phase-curves of other close-in planets, including USP planets (e.g. \citealp{demory2016,morello2019,zieba2022}). The two mathematical definitions of equilibrium temperature differ by a factor of $\sim$1.28. Dividing the temperatures from Table \ref{tab:final_params} by this factor, we recover the previously published values within 1$\sigma$.

Most datasets are shared between this work and that of \cite{giacalone2022}. We performed almost simultaneous analyses using different validation methods, as well as different data detrending techniques, transit light curve parametrisations, and software modelling tools. The 1$\sigma$ consistency confirms the reliability of both sets of results for each of the TOI-1442 and TOI-2445 systems.

In addition, we analysed new high-resolution spectra obtained with Subaru/IRD. The IRD datasets enabled us to set more precise stellar parameters than those reported in the previous literature. In particular, we could shrink the error bars on the stellar masses by a factor of 2--3, on the radii by a factor of 1.1--1.3, and on the effective temperatures by a factor of 1.6. The planetary parameter error bars are comparable between the two studies,  probably being compensated by the use of GPs and/or broader Bayesian priors in our analysis.

\subsection{Comparison with other USP planets}
Figure \ref{fig:P_vs_Rp} shows the radius versus orbital period distribution of the known USP planets. TOI-1442 b and TOI-2445 b are among the 12 validated USP planets with the shortest orbital periods, and likewise their radii are smaller than $2 R_{\oplus}$. If we consider the known sample of 21 USP planets around M dwarfs, TOI-1442 b and TOI-2445 b have the third and the fifth shortest periods, respectively. They also have the seventh and eighth highest equilibrium temperatures. All the USP planets around M dwarfs have radii smaller than $2 R_{\oplus}$, except K2-22 b \citep{sanchis-ojeda2015}. The mass upper limits of $M_{\mathrm{p}} < 8 \, M_{\oplus}$ for TOI-1442 b and $M_{\mathrm{p}} < 20 \, M_{\oplus}$ for TOI-2445 b confirm the sub-giant nature. 
More RV measurements are desirable to place significant constraints on their masses and mean densities, hence their chemical compositions.

We note that the standard deviation in the RV datasets are 2.6--3.8 times larger than the respective mean error bars.
The dispersion in our RV measurements could be due to instrumental systematic offsets, stellar activity, or the possible presence of additional non-transiting planets. Although only a small fraction of USP planets have been detected in multiplanet systems, at least 6 out of the 19 USP planets previously reported around M dwarfs are members of multiplanet systems. These systems are Kepler-32 \citep{fabrycky2012}, Kepler-42 \citep{muirhead2012}, Kepler-732 \citep{morton2016}, LTT-3780 \citep{cloutier2020,nowak2020}, LP 791-18 \citep{crossfield2019}, and LHS-1678 \citep{silverstein2022}.
Another peculiarity of these six systems is that their orbits are aligned so that the outer planets are also transiting. We could also add TOI-1238 to this group; it is a K7-M0 dwarf with two confirmed transiting planets, one of which has an USP \citep{gonzalez-alvarez2021}. For TOI-1238 and LHS-1678, the discovery papers also reported evidence of a non-transiting companion, likely a giant planet or a brown dwarf, in a wide orbit with a period of years. Among the other M dwarf hosts of a transiting USP planet, non-transiting planet candidates have been identified from RV measurements of TOI-1685 \citep{bluhm2021,hirano2021}, TOI-1634 \citep{hirano2021,luque2022}, and GJ-1252 \citep{luque2022}.
The new RV measurements will also be useful to assess the architecture of the planetary systems around TOI-1442 and TOI-2445, which is important to validate formation theories for USP planets around M dwarfs and differences with those around later-type stars (e.g. \citealp{petrovich2020}).

Both TOI-1442 b and TOI-2445 b are suitable  targets to observe their thermal emission spectra. We estimated their emission spectroscopy metric (ESM) to be $9.0_{-1.0}^{+1.1}$ and $11.1_{-1.5}^{+1.7}$, according to the definition given by \cite{kempton2018}. Given their ESM$>$7.5, both TOI-1442 b and TOI-2445 b should be among the top 20 terrestrial targets to be observed in eclipse with the James Webb Space Telescope (JWST)/Mid-InfraRed Instrument (MIRI). We note that the ESM is an estimate of the S/N on the white light eclipse as it would be observed with JWST/MIRI.
Such observations can clarify whether these USP planets are bare rocks stripped of their primordial atmospheres, or whether they have retained substantial gaseous envelopes, and can help us characterise their surface and gas composition.

\section{Conclusions}
\label{sec:conclusions}
We validate the planetary nature of TOI-1442 b and TOI-2445 b, two USP planets with M dwarf stellar hosts. TOI-1442 b has an orbital period of $P = 0.4090682 \pm 0.0000004 \, d$, a radius of $R_{\mathrm{p}} = 1.15 \pm 0.06 \, R_{\oplus}$, and an equilibrium temperature of $T_{\mathrm{p,eq}} = 1357_{-42}^{+49} \, K$. TOI-2445 b has an orbital period of $P = 0.3711286 \pm 0.0000004 \, d$, a radius of $R_{\mathrm{p}} = 1.33 \pm 0.09 \, R_{\oplus}$, and equilibrium temperature of $T_{\mathrm{p,eq}} = 1330_{-56}^{+61} \, K$.
We report $3\sigma$ upper limits on their masses of $M_{\mathrm{p}} < 8 M_{\oplus}$ and $M_{\mathrm{p}} < 18 M_{\oplus}$, respectively. The upper mass limits are obtained by assuming a pure iron composition.
We also provide precise stellar parameters from previously unpublished high-resolution spectra.

It would be interesting to follow-up on these targets with high-precision RV facilities to improve their planetary mass measurements (to constrain their bulk compositions) and possibly detect other planetary companions. They are also suitable targets for emission spectroscopy with JWST.

\begin{acknowledgements}
G. M. has received funding from the European Union's Horizon 2020 research and innovation programme under the Marie Sk\l{}odowska-Curie grant agreement No. 895525, and the Ariel Postdoctoral Fellowship Program. This work is partly financed by the Spanish Ministry of Economics and Competitiveness through grants PGC2018-098153-B-C31.
This work was partly supported by MEXT/JSPS KAKENHI Grant Numbers JP17H04574, JP18H05439, JP20J21872, JP18H05439, JP20K14521, JP20K14518, JP21K20376, JP21K13975, JP18H05439, JP18H05442, JP15H02063, JP22000005, JP18H05439, JP19K14783, JP21H00035, JP21K20388, JP21K13955, JST CREST Grant Number JPMJCR1761, Astrobiology Center SATELLITE Research project AB022006, and Astrobiology Center PROJECT Research AB031014.

R.L. acknowledges funding from University of La Laguna through the Margarita Salas Fellowship from the Spanish Ministry of Universities ref. UNI/551/2021-May 26, and under the EU Next Generation funds.

J. K. gratefully acknowledges the support of Swedish National Space Agency (SNSA; DNR 2020-00104) and of the Swedish Research Council  (VR: Etableringsbidrag 2017-04945)

M.S. acknowledges the support of the Italian National Institute of Astrophysics (INAF) through the project ``The HOT-ATMOS Project: characterizing the atmospheres of hot giant planets as a key to understand the exoplanet diversity'' (1.05.01.85.04).

This paper is based on data collected at the Subaru Telescope, which is located atop Maunakea and operated by the National Astronomical Observatory of Japan (NAOJ). We wish to recognize and acknowledge the very significant cultural role and reverence that the summit of Maunakea has always had within the indigenous Hawaiian community. A part of our data analysis was carried out on common use data analysis computer system at the Astronomy Data Center, ADC, of the National Astronomical Observatory of Japan.
The research leading to these results has received funding from the ARC grant for Concerted Research Actions, financed by the Wallonia-Brussels Federation, and from the French Community of Belgium in the context of the FRIA Doctoral Grant awarded to MT. TRAPPIST is funded by the Belgian Fund for Scientific Research (Fond National de la Recherche Scientifique, FNRS) under the grant PDR T.0120.21.  MG and EJ are F.R.S.-FNRS Senior Research Associate.

This work makes use of observations from the LCOGT network. Part of the LCOGT telescope time was granted by NOIRLab through the Mid-Scale Innovations Program (MSIP). MSIP is funded by NSF.

This paper is based on observations made with the MuSCAT3 instrument, developed by the Astrobiology Center and under financial supports by JSPS KAKENHI (JP18H05439) and JST PRESTO (JPMJPR1775), at Faulkes Telescope North on Maui, HI, operated by the Las Cumbres Observatory.

\end{acknowledgements}

%
%
\clearpage \clearpage

\bibliographystyle{aa} 
\bibliography{biblio} 

\clearpage

\end{document}